\documentclass[preprint]{aastex}
\usepackage{amsmath}

\shorttitle{THE MASS TRANSFER IN CATACLYSMIC VARIABLES}
\shortauthors{Sirotkin \& Kim}

\slugcomment{Accepted for publication in the ApJ}

\newcommand\simlt{\lower2pt
   \hbox{$\;\buildrel{\scriptstyle <}\over{\scriptstyle\sim} \; $}}
\newcommand\simgt{\lower2pt
   \hbox{$\;\buildrel{\scriptstyle >}\over{\scriptstyle\sim} \; $}}

\newcommand {\Aise}{\alpha_{\rm isen}}

\newcommand {\tG} {{\tau_{\rm \rm G}}}

\newcommand {\tJ} {{\tau_{\rm \rm J}}}
\newcommand {\tM} {{\tau_{\rm M}}}

\newcommand {\MHBM} {{0.075\rm\;M_\odot}}

\newcommand {\tKH} {{\tau_{\rm KH}}}

\newcommand {\Pgam} {{P_{\rm gap}^{~-}}}
\newcommand {\Pgap} {{P_{\rm gap}^{~+}}}
\newcommand {\agam} {{\alpha_{\rm gap}^{-}}}
\newcommand {\agap} {{\alpha_{\rm gap}^{+}}}
\newcommand {\Pmin} {{P_{\rm min}}}
\newcommand {\Pt} {{P_{\rm turn}}}

\newcommand {\rhob} {{\bar{\rho}}}

\newcommand {\Ln} {{L_{\rm nuc}}}
\newcommand {\LnO} {{L_{\rm nuc,0}}}

\newcommand {\Lg} {{L_{\rm g}}}

\newcommand {\Rs} {{\rm \;R_\odot}}

\newcommand {\Ms} {{\rm \;M_\odot}}

\newcommand {\adMwd}{{\langle\dot{M}_1\rangle}}
\newcommand {\dMwd}{{\dot{M}_1}}

\newcommand {\dM}{{\dot{M}_2}}
\newcommand {\dMg}{{\dot{M}_{2,\rm{R}}}}

\newcommand {\xS}{{\xi_{\rm S}}}

\newcommand {\xE}{{\xi_{\rm E}}}

\newcommand {\xR}{{\xi_{\rm R}}}

\newcommand {\dMd}{{\rm\; M_\odot \;\rm yr^{-1}}}

\newcommand {\bg} {{\bar{g}}}

\begin{document}

\title{A SEMI-EMPIRICAL MASS-LOSS RATE IN SHORT-PERIOD CATACLYSMIC VARIABLES}

\author{Fedir V.\ Sirotkin \& Woong-Tae Kim}

\email{sirotkin.f.v@gmail.com, wkim@astro.snu.ac.kr}

\affil{Department of Physics and Astronomy, FPRD, Seoul National
University, Seoul, 151-742, South Korea}

\begin{abstract}
The mass-loss rate of donor stars in cataclysmic variables (CVs) is
of paramount importance in the evolution of short-period CVs.
Observed donors are oversized in comparison with those of isolated
single stars of the same mass, which is thought to be a consequence
of the mass loss. Using the empirical mass-radius relation of CVs
and the homologous approximation for changes in effective
temperature $T_2$, orbital period $P$, and luminosity of the donor
with the stellar radius, we find the semi-empirical mass-loss rate
$\dM$ of CVs as a function of $P$. The derived $\dM$ is at
$\sim10^{-9.5}$--$10^{-10}$ $\dMd$ and depends weakly on $P$ when
$P>90$ min, while it declines very rapidly towards the minimum
period when $P<90$ min, emulating the $P$--$T_2$ relation. Due to
strong deviation from thermal equilibrium caused by the mass loss,
the semi-empirical $\dM$ is significantly different from, and has a
less-pronounced turnaround behavior with $P$ than suggested by
previous numerical models. The semi-empirical $P$--$\dM$ relation is
consistent with the angular momentum loss due to gravitational wave
emission, and strongly suggests that CV secondaries with $0.075\Ms <
M_2 <0.2\Ms$ are less than 2 Gyrs old. When applied to selected
eclipsing CVs, our semi-empirical mass-loss rates are in good
agreement with the accretion rates derived from the effective
temperatures $T_1$ of white dwarfs, suggesting that $\dM$ can be
used to reliably infer $T_2$ from $T_1$. Based on the semi-empirical
$\dM$, SDSS 1501 and 1433 systems that were previously identified as
post-bounce CVs have yet to reach the minimal period.
\end{abstract}


\section{Introduction}\label{sec:intro}

\par
Cataclysmic variables (CVs) are semi-detached binaries with a white
dwarf (WD) primary and a low-mass companion which transfers mass to
a WD through Roche-lobe overflow. The observed orbital period
distribution of CVs has three important features: a sharp short
period cut-off at the orbital period of $\Pmin\simeq76.2$ min,
referred to as the ``minimum period'', the period gap,  a dearth of
systems with periods between $\Pgam\simeq129$ min and
$\Pgap\simeq192$ min \citep{c_Knigge_2006}, and the ``period minimum
spike'', a significant accumulation of systems at $P\sim 80$--$86$
min \citep{c_Gansike_2009}. For CVs with period above $\Pgap$, the
dominant angular momentum loss mechanism is thought to be the
magnetic braking, while the gravitational radiation may be
responsible for the angular momentum loss for CVs below $\Pgam$. In
this work, we focus on CVs only with $P<\Pgam$.

\par
A characteristic feature of CVs is the mass transfer from a
secondary to a WD. Since the gas stream carries substantial angular
momentum, it forms an accretion disk around the WD. The reaction of
a CV on the mass transfer can be qualitatively understood with the
help of the dimensionless parameter
\begin{equation}\label{eq_t_def}
 \tau=\frac{\tM}{\tKH},
\end{equation}
where $\tM$ and $\tKH$ denote the mass-loss and Kelvin-Helmholtz
timescales of the donor, respectively \citep{c_Pacz_1981}. At the
initial phase of mass transfer, $\tau \gg 1$, so that the donor
remains close to thermal equilibrium and thus follows the
mass-radius relation (MRR) of main-sequence stars
\begin{equation}
R_2 \propto M_2^\xE,
\end{equation}
with the effective mass-radius exponent $\xE$ close to
$0.64^{+0.02}_{-0.02}$ \citep{c_Knigge_2006}. Here, the subscript
``2'' denotes the quantities of CV donors. Both $R_2$ and $P$ of the
CV decrease as the donor keeps losing mass during this early phase
of CV evolution.

\par
As the mass transfer continues, the binary shrinks due to the
angular momentum loss, reducing $\tau$. When $\tM$ and $\tKH$ become
comparable to each other ($\tau\approx1$), the donor becomes out of
thermal equilibrium. Since the surface luminosity is no longer in
balance with the nuclear luminosity, the donor expands, causing
$\xE$ to decrease with time \citep{c_King_1988}. Using the Hayashi
theory and homologous approximation, \citet{c_Stehle_1996} studied
the reaction of the donor to mass loss and found that $R_2$
increases with decreasing $\tau$. Observations confirm this
theoretical prediction, showing that the radii of donors in
short-period CVs are $\sim5$--$50\%$ larger than those of isolated
main-sequence stars \citep{c_Patterson_2005, c_Knigge_2006}.

\par
The effective mass-radius exponent $\xE$  becomes smaller as $M_2$
decreases towards the brawn dwarfs masses. The change in the MRR of
donors slows down the temporal change in the orbital period as the
donor continues losing mass. The orbital period reaches a minimum
value when $\xE\approx1/3$ at which $dP/dM_2=0$
\citep{c_Rappaport_1982}. As $\tau$ decreases further, $\xE$
declines to $0.21^{+0.05}_{-0.10}$ \citep{c_Knigge_2006}. This in
turn leads to an increase in $P$. Such a turnaround of the orbital
period at $\Pt$ is referred to as the ``period bounce''.

\par
The evolutionary scenario described above suggests that the
mass-loss rates $\dM$ (and the related thermal relaxation) of donors
control the evolution of CVs.  It also implies that the observed
distribution of CVs versus the orbital period may be a consequence
of a certain relation between $\dM$ and $P$. Since $\dM(P)$ not only
provides a way to estimate the ages of CVs but also can be used to
check the dominant mechanism for angular momentum loss, finding a
reliable $P$--$\dM$ relationship is of great importance  in
understanding CV evolution.

\par
There have been numerous studies to derive the $P$--$\dM$ relation;
these can be categorized into two groups: purely theoretical work
(e.g., \citealt{c_Rappaport_1982,c_Ritter_1988,c_King_1988,
c_Kolb_1999, c_Howell_2001}) and semi-empirical work (e.g.,
\citealt{c_Patterson_1984, c_Urban_2006, c_Townsley_2009}). The
first, theoretical studies made use of stellar evolution codes
combined with the analytical mass-transfer models of
\citet{c_Ritter_1988} or others. The common results of these
numerical experiments are that (1) the orbital period exhibits the
turnaround behavior at $\Pt$ that is about $\sim10\%$ smaller than
the observed $\Pmin$ \citep{c_Rappaport_1982, c_Kolb_1999,
c_Howell_2001} and (2) compared to isolated stars, the secondaries
in short-period CVs are oversized due to the mass loss but only by
$\sim5$--$20\%$, which is about a factor $\sim1.5$--$2$ smaller than
the observed CV sizes. Since the bloating of the donor depends on
the efficiency of the mass loss, these results suggest either the
numerical models underestimate $\dM$ above $\Pt$, or the efficiency
of the angular momentum loss is higher than expected
\citep{c_Kolb_1999, c_Renvoize_2002}. Despite the attempts to
include the effects of tidal and rotational perturbations
\citep{c_Rezzolla_2001, c_Renvoize_2002, c_Kolb_1999}, additional
angular momentum loss via magnetic braking \citep{c_Andronov_2003},
accretion disk winds and magnetic propeller (see
\citealt{c_Barker_2003} for review), etc., clear explanations for
the discrepancies between $\Pt$ and $\Pmin$ and between the
empirical and numerical MRRs have yet to be found.

\par The second, semi-empirical approach uses, for instance, the effective temperatures of WDs
\citep{c_Townsley_2003, c_Townsley_2004,c_Townsley_2009} or
accretion disk luminosities \citep{c_Patterson_1984} to infer the
accretion rates $\dMwd$. The physical basis for the former is that
WDs in CVs are observed to be hotter than expected from their ages
\citep{c_Sion_1995}, and such ``overheating" of WDs may be a result
of the compressional heating due to the gas piled up on their
surfaces \citep{c_Townsley_2003}. For the latter method, the
accretion luminosities can give direct information on $\dMwd$ if the
disk properties are well constrained. When applied to dwarf novae
(DNe) that accrete sporadically and have thermally unstable disks,
$\dMwd$ calculated from the two methods have a different scatter for
given $P$. The mass-transfer rates based on the accretion
luminosities probably measure an accretion rate averaged over short
time intervals, e.g., a few decades, which makes them quite
sensitive to short-time disk variability \citep{c_Townsley_2009}.
This inevitably leads to a high dispersion in $\dMwd$ estimated from
the disk luminosities. On the other hand, the effective temperatures
of WDs are likely to give a relatively long-term averaged $\dMwd$,
although this method may suffer from uncertainties related to
unknown properties such as masses and radii of WDs as well as some
auxiliary assumptions on their structures and boundary layers
\citep{c_Patterson_2008}.

\par
In this paper we construct another semi-empirical relation between
the mass-loss rate and the orbital period, and compare the result
with those of the purely theoretical and the semi-empirical
approaches mentioned above. We start from the empirical MRR of
\citet{c_Knigge_2006} for superhumping CVs, and apply the homologous
approximation in order to calculate the dependencies of the
effective temperature and gravothermal luminosity upon the bloating
of the donor. Assuming that the observed MRR of CV donors is a
consequence of the mass loss and thermal relaxation processes, we
calculate the mass-loss rates. We will show that the resulting
mass-loss rates are broadly in line with those on the basis of the
effective temperatures of WDs, while different from those of the
purely theoretical models of \citet{c_Kolb_1999}. We will also show
that our results are consistent with the angular momentum loss of
CVs due primarily to the emission of gravitational waves. Since
radii of CV donors likely trace the long-term evolution of $\dM$
\citep{c_Knigge_2006} and depend only weakly on details of the
mass-transfer processes, the resulting mass-loss rates from the
empirical MRR are less model-dependent and less sensitive to the
short-term effects than the other semi-empirical estimates that rely
on accretion activities.

\par The organization of this paper is as follows.
In \S\ref{sec_P_Te}, we introduce the theoretical MRR for isolated
single stars and the empirical MRR for CVs donors, and define the
bloating factor. We then describe a method to find the effective
temperature and gravothermal luminosity of donors by using the
bloating factor. In \S\ref{sec_dM}, we calculate the semi-empirical
mass-loss rate, construct the mass-loss history to estimate the CV
ages, and show that the derived mass-loss rate is consistent with CV
evolution by angular momentum loss via the emission of gravitational
waves. In \S\ref{sec_Dis}, we apply our results to selected
eclipsing short-period CVs and show that the semi-empirical
mass-loss rate is consistent with the accretion rates inferred from
the effective temperatures of WDs. We also suggest a way to estimate
the effective temperatures of donors assuming that mass is
conserved. Finally, we summarize our results in \S\ref{sec_Con}.

\section{Mass-radius and Period-effective Temperature Relations}\label{sec_P_Te}

\par We consider a CV consisting of a WD (or accretor) with mass $M_1$
and a secondary (or donor) with mass $M_2$, radius $R_2$, and
effective temperature $T_2$. We assume that the donor fills its
critical equipotential surface. In accordance with the results of
\citet{c_Patterson_2005} and \citet{c_Knigge_2006}, we assume
$M_1=0.75\Ms$ throughout this paper. We use an additional subscript
``0'' to denote the quantities of isolated, single stars in
hydrostatic and thermal equilibrium to which CV secondaries will be
compared. In this section, we use the empirical relationship between
masses and radii of CV secondaries to calculate the bloating factor
in comparison with the single stars of the same mass. We then treat
the bloating factor as perturbations to the properties of single
stars in order to find the orbital period, effective temperature,
and gravothermal luminosity of CVs.

\subsection{Mass-radius Relations}\label{ssec_MRR}

\par As the MRRs of single stars that we use as the references to those of
CV secondaries, we take \emph{composite} relations constructed in a
manner similar to \citet{c_Knigge_2006}: the BCAH98 isochrone of
\citet{c_Baraffe_1998} for stars with $0.075\Ms\leq M_2<0.3\Ms$, the
DUSTY isochrone of \citet{c_Baraffe_2002} for $0.06\Ms\leq
M_2<0.075\Ms$ ($1500\lesssim T_2\lesssim2000$ K), and the COND
isochrone of \citet{c_Baraffe_2003} for $M_2<0.06\Ms$.

\par
\begin{figure}
 \epsscale{0.9}
 \plotone{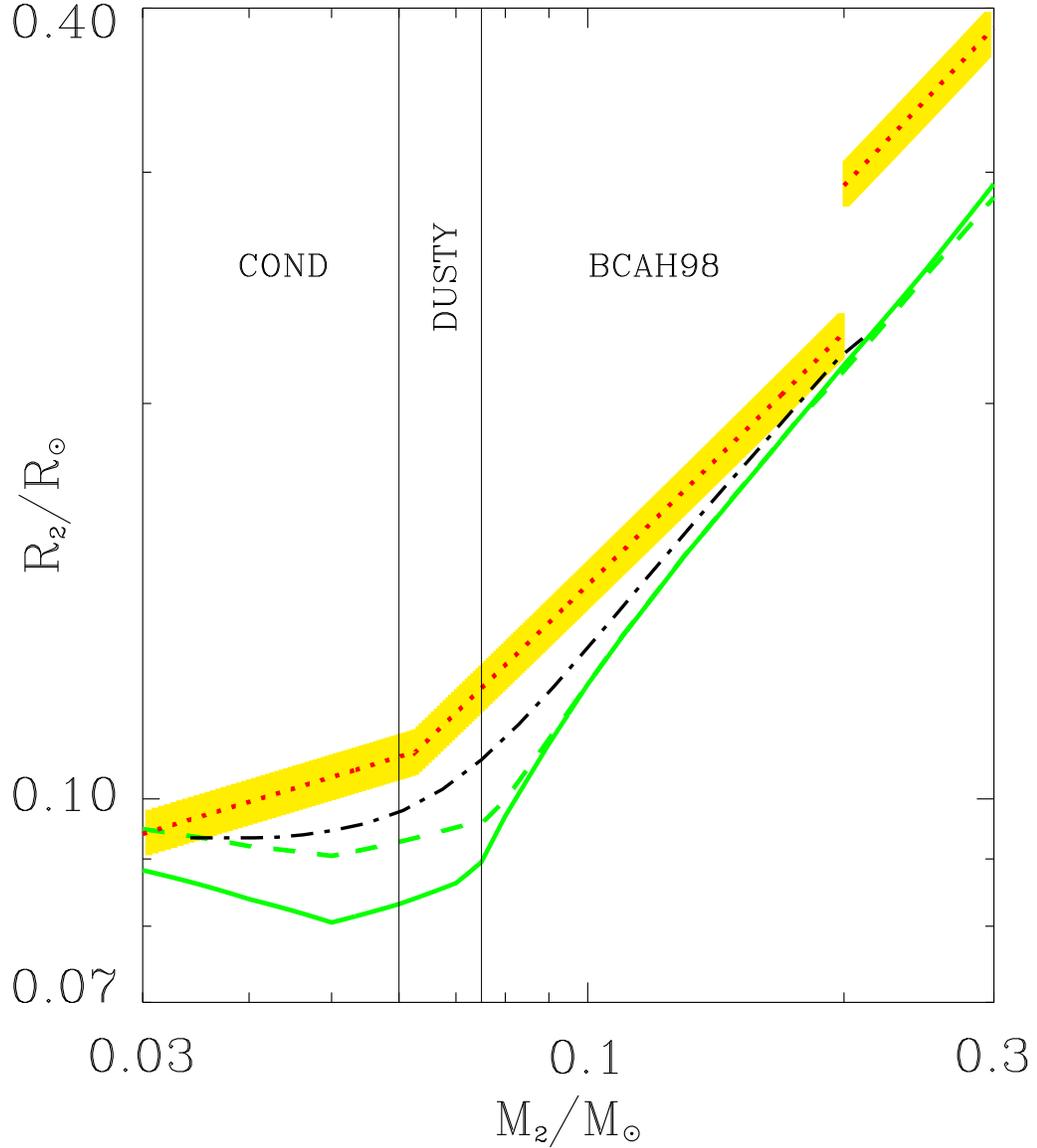}
 \caption{
Mass-radius relations of CV donor stars. The dotted line plots the
empirical MRR from \citet{c_Knigge_2006} with the 4\% uncertainties
indicated by the shaded region. The dashed and solid lines represent
the unperturbed MRR of singles stars for the B1 and B5 sequences,
respectively. The dot-dashed line is for the MRR of evolving donors
from the numerical models of \citet{c_Kolb_1999}. The thin vertical
lines mark the boundaries of the adopted isochrones: BCAH98 for
$0.075\leq M_2<0.3\Ms$; DUSTY for $0.06\leq M_2<0.075\Ms$; COND for
$M_2<0.06\Ms$.
 }\label{pic_m_r}
\end{figure}

 There are quite large uncertainties in the estimation of CV ages.
Period bouncers are thought of as a final stage of CV evolution,
suggesting that they are older than their ``parent'' CVs (see, e.g.,
\citealt{c_Pacz_1981,c_Rappaport_1982, c_Kolb_1999}). Although
numerical experiments predict that CVs cross the period gap at
$\sim1$ Gyr and reach $\Pt$ in $\sim4$ Gyr after the onset of the
mass loss below the gap (e.g., \citealt{c_Kolb_1999,
c_Howell_2001}), temporal evolution of CVs depends strongly on the
mass-loss rate that is not well constrained observationally. To
cover a wide range of CV ages, we adopt two composite isochrones
with 1-Gyr and 5-Gyr ages (hereafter B1 and B5 sequences,
respectively)\footnote{The composite isochrones we adopt are
slightly different from those in \citet{c_Knigge_2006} who took the
5-Gyr isochrone for donors with $M\gtrsim0.075\Ms$ and 1-Gyr
isochrone for $M<0.075\Ms$. The choice of the 1-Gy isochrone in
\citet{c_Knigge_2006} was to represent the numerical results of
\citet{c_Kolb_1999}.}. Figure \ref{pic_m_r} plots the composite
$R_{2,0}$ of single stars as functions of $M_2$ for the B1 and B5
sequences, as dashed and solid lines, respectively. Both sequences
give almost the same radii above $\MHBM$, while below this point the
stellar radii in the B1 sequence are about $\sim10\%$ larger than in
the B5 sequence.

\par
As the MRR of CV secondaries, we take the relation derived by
\citet{c_Knigge_2006} from the observed properties of superhumping
CVs:
\begin{equation}\label{eq_Kn_mr_fit}
 \frac{R_2}{\Rs}=
 \begin{cases}
 0.110\bigl(\frac{M_2}{0.063\Ms}\bigr)^{0.21},&\;\; 0.063\Ms<M_2,\\
 0.230\bigl(\frac{M_2}{0.2\Ms}\bigr)^{0.64},&\;\;  0.063\Ms\leq M_2<0.2\Ms, \\
 0.299\bigl(\frac{M_2}{0.2\Ms}\bigr)^{0.67},&\;\;  0.2\Ms\leq
 M_2.
 \end{cases}
\end{equation}
This empirical fit assumes $M_1=0.75\Ms$ and incorporates itself the
additional empirical constraints that reproduce the observed
locations of the period gap and the period minimum, such that
$\Pgam$ occurs at $M_2=0.2\pm{0.02}\Ms$ and $\Pmin$ at
$M_2=0.063\pm{0.009}\Ms$. The intrinsic errors in $R_2$ for this
empirical MRR are estimated to be about $2$--$3\%$
\citep{c_Knigge_2006}; we take allowance for $4\%$ uncertainties in
$R_2$  throughout this work. We note a caveat that our choice of the errors may underestimate real values below $\MHBM$ where only a few systems are observed and thus there is practically no calibrator. In addition, the masses of WDs may differ from the
assumed mass of $0.75\Ms$, which also can change $R_2$. Equation
(\ref{eq_Kn_mr_fit}) is plotted in Figure \ref{pic_m_r} as dotted
line with the shaded region denoting the associated uncertainties in
$R_2$. Columns (1) and (2) in Table \ref{ta_seq} give the
$M_2$-$R_2$ relation.

For comparison, Figure \ref{pic_m_r} also plots as a dot-dashed line
the MRR of CVs with $0.03\Ms<M_2<0.21\Ms$ from the numerical models
of \citet{c_Kolb_1999}, which clearly has smaller radii than the
empirical relation, typically by $\sim5$--$10\%$. This indicates
that the deviation of real donors from thermal equilibrium is larger
than that expected from the numerical models.

%fig1

\subsection{Bloating Factor}\label{ssec_al}

\par
\begin{figure}
 \epsscale{0.9}
 \plotone{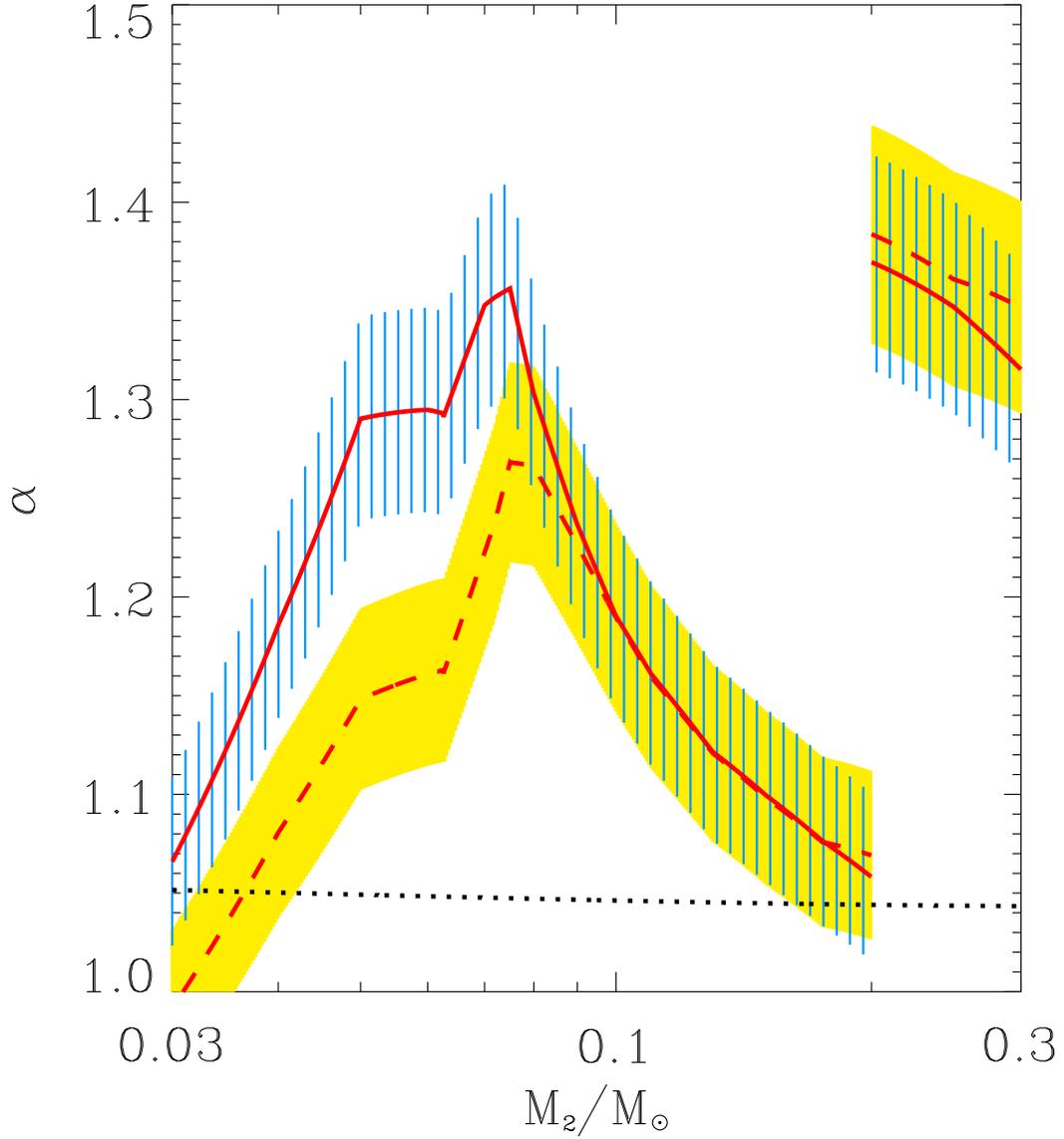}
 \caption{
Bloating factors $\alpha$ of CV donors as functions of their mass
$M_2$. The dashed and solid lines are for the B1 and B5 sequences,
respectively, with the shaded and hatched regions corresponding to
the uncertainties. The dotted line shows the bloating factor of
Roche-lobe filling polytropes with $n=3/2$ due to the isentropic
expansion caused by tidal and rotational perturbations, when its
companion has a mass of $M_1=0.75\Ms$.
 }\label{pic_m_al}
\end{figure}

 We define the bloating factor
\begin{equation}\label{eq_alpha_def}
 \alpha=\frac{R_2}{R_{2,0}},
\end{equation}
as the relative size of a CV secondary to a single star with the
same mass. Figure \ref{pic_m_al} plots $\alpha$ as dashed and solid
lines for the B1 and B5 sequences, respectively, with the shaded and
hatched regions representing the corresponding uncertainties. The
discontinuity of $\alpha$ at $M_2=0.2\Ms$ corresponds to the period
gap. Since $P\propto \alpha^{3/2}$ (see below), the ratio of the
bloating factors above ($\agap$) and below ($\agam$) the gap
satisfies $\agap/\agam=(\Pgap/\Pgam)^{2/3} = 1.3$, in good agreement
with the width of the gap. At the upper edge of the gap, the donors
detach from their critical lobes and the mass loss stops. Since the
donors are thought to be oversized due to the mass loss, they shrink
within the gap. When the donors come out the lower edge of the gap,
they resume the mass loss and become oversized again.

%fig2

\par Figure \ref{pic_m_al} shows that the bloating factor increases with
the decreasing donor mass for $M_2>\MHBM$ and then decreases with
decreasing $M_2$ below this point. This can be explained
qualitatively as follows. Since stars with $M_2<0.3\Ms$ are fully
convective and isentropic, they are well described by polytropes
with index $n=3/2$ \citep{c_Burrows_1993}. The MRR for these
polytropes
\begin{equation}\label{eq_r_m_polytrope}
 R_2\propto K M_2^{-1/3},
\end{equation}
with $K$ being the pressure constant \citep{c_Chandra_1939}, then
gives
\begin{equation}\label{eq_alpha_def_K}
 \alpha=\frac{K}{K_0}.
\end{equation}
The mass loss is capable of driving the donor out of thermal
equilibrium. This raises the gravothermal luminosity, $\Lg=L_2-\Ln$,
where $\Ln$ is the nuclear luminosity of the donor and $L_2$ is the
surface luminosity. For fully convective stars, the change of $K$
depends on $\Lg$ and $\tKH$ as
\begin{equation}\label{eq_dk_K}
 \frac{\dot{K}}{K}\propto-\frac{1}{\tKH}\frac{\Lg}{L_2},
\end{equation}
\citep{c_King_1988, c_Ritter_1996}. Combining equations
(\ref{eq_r_m_polytrope}), (\ref{eq_alpha_def_K}), and
(\ref{eq_dk_K}), one obtains
\begin{equation}\label{eq_da_a}
 \frac{\dot{\alpha}}{\alpha}\propto-\frac{1}{\tKH}
 \frac{\Lg}{L_2}+\frac{1}{\tM}\biggl(\xE_0+\frac{1}{3}\biggr),
\end{equation}
where $\xE_0$ is the mass-radius exponent of unperturbed, singe
stars. At the lower edge of the period gap, a donor remains close to
thermal equilibrium so that $\Lg\approx 0$ and $\xE_0\approx1$. In
this case, the second term in the right-hand side of equation
(\ref{eq_da_a}) dominates and $\dot{\alpha}/\alpha>0$, indicating
that $\alpha$ increases as the mass loss continues. When the donor
reaches the minimum hydrogen burning mass (MHBM) $\approx\MHBM$, the
nuclear burning stops and donors are born as brown dwarfs, so that
$\Lg/L_2=1$. In the brown dwarf regime, on the other hand, the
internal pressure is dominated by the degeneracy pressure with $K$
depending mainly on the chemical composition \citep{c_Burrows_1993}.
In this case, $\xE_0$ tends to $-1/3$ and the first term in the
right-hand side of equation (\ref{eq_da_a}) dominates, resulting in
$\dot{\alpha}/\alpha<0$. For brown-dwarf donors, the change in the
equation of state partially compensates for the increasing tendency
of the bloating factor due to the mass loss.

\par In addition to the mass loss, tidal and rotational perturbations
can also change the donor size, depending on the polytropic index
and the mass ratio. We assume that the applied perturbations do not
change the specific entropy, which is reasonable since the stellar
distortions occur in dynamical timescales.  In Appendix
\ref{ap_isentropic}, we calculate the enlargement factor $\Aise$ of
a Roche-lobe filling donor as a function of $n$ for $q=0.1$. The
resulting $\Aise$ for $n=3/2$ polytropes against $M_2$ is plotted in
Figure \ref{pic_m_al} as a dotted line. Note that the expansion of
low-mass donors due to the tidal and rotational distortions is
$\sim4$--$5\%$, insensitive to the mass ratio for $q\sim0.04$--$0.4$
(see also, e.g., \citealt{c_Uryu_1999, c_Renvoize_2002,
c_SWTK_2009}). It can be seen that while $\Aise$ is, in general,
considerably smaller than the observed bloating factor for stars
with $0.04\Ms\simlt M_2\simlt0.15\Ms$, it is comparable to $\alpha$
when $M_2\approx0.2\Ms$ corresponding to $\Pgam$ or when
$M_2<0.04\Ms$ as donors approach $\Pmin$.  This indicates that the
effect of the isentropic expansion of donors on the bloating factor
is non-negligible for CVs that just resume mass loss or are below
the minimum period.

\subsection{Orbital Period and Effective Temperature}\label{ssec_Te}

\par
For a given MRR, the orbital period can be determined by the Roche
lobe geometry and Kepler's law as
\begin{equation}\label{eq_mean_den_law}
 P^2=\frac{3\pi}{G\rhob}\frac{q}{f^3(1+q)},
\end{equation}
where $\rhob={3M_2}/{4\pi R_2^3}$ is the mean density of the donor,
$f=(2/3^{4/3})q^{1/3}/(1+q)^{1/3}$ is the relative size of the donor
to the orbital separation, and $q=M_2/M_1$ is the mass ratio (e.g.,
\citealt{c_Kopal_1972}). Equation (\ref{eq_mean_den_law}) implies
that the orbital period depends on the bloating of the donor as
\begin{equation}\label{eq_P_cor}
P=P_0\alpha^{3/2}.
\end{equation}
Clearly, the orbital period increases with the radius of a
secondary.

\par
Since donors below the period gap are fully convective, their
luminosities are determined by the conditions at the very outermost
layers \citep{c_Kipp_1990}, so that one can neglect the influence of
the distortions on the central parts of a donor. To obtain the
properties of the photospheric layers, we use the standard Hayashi
theory. For the opacity law of the form
\begin{equation}
 \kappa=\kappa_0P_g^aT^b,
\end{equation}
with the gas pressure $P_g$, the ratio of the effective temperatures
satisfies
\begin{equation}\label{eq_Te_cor_KW}
 \frac{T_2}{T_{2,0}}= \alpha^\gamma,
\end{equation}
where $\gamma=(3a-1)/(5a+2b+5)$ (e.g, \citealt{c_Kipp_1990}, see
also \citealt{c_Stehle_1996}). For the realistic opacity laws
applicable for low-mass main-sequence stars, $\gamma$ has a very
small value. If the opacity is governed by H$^{-}$, for example,
$a\approx0.5$ and $b\approx4.5$, leading to $\gamma\approx 0.03$. If
the molecular absorption instead is a dominant opacity source
($a\approx1$ and $b\approx0$), $\gamma\approx0.2$. Therefore, donors
in short-period CVs have approximately the same effective
temperatures as the corresponding isolated single stars of identical
masses (e.g., \citealt{c_Stehle_1996}).

\par
\begin{figure}
 \epsscale{0.9}
 \plotone{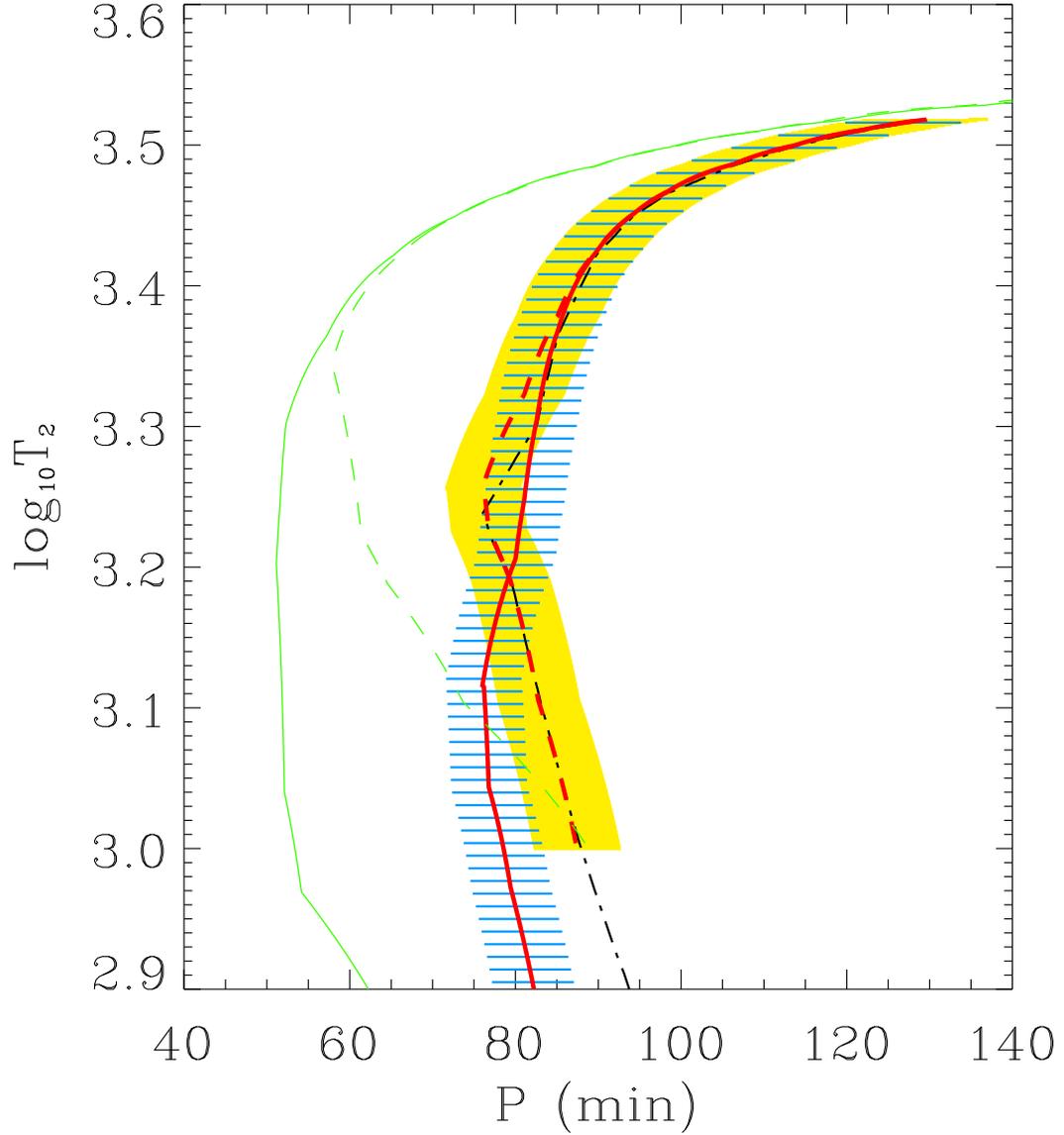}
 \caption{
Dependence of the effective temperature $T_2$ of a donor on the
orbital period $P$. The thick dashed and solid lines plot the
semi-empirical $P$--$T_2$ relations for the B1 and B5 sequences,
respectively, with the dashed and hatched regions denoting the
corresponding uncertainties. The thin dashed and solid lines give
the unperturbed $P_0$--$T_{2,0}$  relations for the B1 and B5
sequences, respectively. The semi-empirical $P$--$T_2$ relation of
\citet{c_Knigge_2006} is plotted as the dot-dashed line.
 }\label{pic_Te_P}
\end{figure}

 To construct the semi-empirical $P$--$T_2$ relations, we first use
the theoretical $M_2$--$T_{2,0}$ relation for isolated stars and
calculate $P_0$ from equation (\ref{eq_mean_den_law}) by fixing
$M_1=0.75\Ms$. We then use the empirical MRR and the bloating factor
of donors to calculate $P$ and $T_2$ from equations (\ref{eq_P_cor})
and (\ref{eq_Te_cor_KW}). The $P_0$--$T_{2,0}$ relations are plotted
in Figure \ref{pic_Te_P} as thin lines, while the thick lines are
for the semi-empirical $P$--$T_2$ relations. The dashed and solid
lines correspond to the B1 and B5 sequences, respectively. Again,
the shaded and hatched regions give the corresponding uncertainties.
Columns (3), (6), and (11) in Table \ref{ta_seq} list the values of
$P$ and $T_2$ against $M_2$ for both sequences. For a given mass,
the expansion of a donor increases the orbital period while the
effective temperature keeps almost unchanged. The largest difference
between $P$ and $P_0$ occurs at $T_2\sim2000-2200$ K, corresponding
to $M_2\approx\MHBM$ where the bloating factor is maximal. The
dot-dashed line plots the semi-empirical result of
\citet{c_Knigge_2006}. Because of the difference in the adopted ages
of the isochrones, the result of \citet{c_Knigge_2006} for
$M_2>\MHBM$ (with $T_2>2000$ K) and $M_2<0.06\Ms$ (with $T_2<1800$
K) is in agreement with our results for the B5 and B1 sequences,
respectively. The donors at $\Pmin$ have mass $M_2\sim
0.054-0.072\Ms$, corresponding to $T_2\sim 1000$--$1800$ K and $\sim
1600$--$2100$ K for the B5 and B1 sequences, respectively. The right
choice of the stellar age is quite important in evaluating the
mass-loss rate near $\Pmin$ since it depends rather sensitively on
$T_2$ as $\dM\propto T_2^4$ \citep{c_Rappaport_1982}.

%fig3

\subsection{Luminosities and Timescales}\label{ssec_Lg}

\par
Combining equation (\ref{eq_Te_cor_KW}) with the Stefan-Boltzmann
law
\begin{equation}\label{eq_st_bl}
 L_2=4\pi\sigma R_2^2T_2^4,
\end{equation}
where $\sigma$ is Stefan-Boltzmann constant, one can show that the
change in the stellar luminosity is given by
\begin{equation}\label{eq_Ls_cor}
 L_2/L_{2,0}=\alpha^{\frac{22a+4b+6}{5a+2b+5}}.
\end{equation}
Although the effective temperature does not change much with the
bloating factor, the additional $\alpha^2$ term in the
Stefan-Boltzmann law increases $L_2/L_{2,0}$ significantly.  Columns
(5) and (10) in Table \ref{ta_seq} give the values of $L_2$ for the
B1 and B5 sequences, respectively.

\par
The nuclear luminosity $\Ln$ is determined by the energy produced in
the central parts of a donor, so that $\Ln\simeq M_2\epsilon_c$,
where $\epsilon_c$ is the thermonuclear reaction rate given by
\begin{equation}\label{eq_enrr_fit}
 \epsilon_c\propto T_c^\nu \rho_c^{\eta-1},
\end{equation}
with $T_c$ and $\rho_c$ denoting the central temperature and
density, respectively. Since $T_c \propto K \rho_c^{2/3} \propto \alpha^{-1}$
for a monatomic ideal gas, the change in the nuclear
luminosity due to the bloating factor is
\begin{equation}\label{eq_Ln_cor}
 \frac{\Ln}{\LnO}= \alpha^{3-3\eta-\nu}.
\end{equation}
Theoretical models for stellar structure and evolution give
$\nu\approx6$ and $\eta\approx2$ \citep{c_Burrows_1993} for
$T_c\simeq3\times10^6$ K and $\rho_c\simeq 150$ g cm$^{-3}$, typical
for the centers of low-mass stars.

\par Since $L_{2,0}=\LnO$ for isolated stars in thermal equilibrium,
the gravothermal luminosity satisfies
\begin{equation}\label{eq_Lg_def}
 \frac{\Lg}{L_2}=1-\frac{\Ln}{\LnO}\frac{L_{2,0}}{L_{2}}.
\end{equation}
Thermonuclear burning stops in CV secondaries once they reach the
MHBM $\approx\MHBM$. At this point, donors are effectively born as
brown dwarfs and the gravitational collapse becomes the dominant
source of the stellar radiation. That is, ${\Lg}/{L_2}=1$ for
$M_2\leq\MHBM$.

\par
\begin{figure}
 \epsscale{0.9}
 \plotone{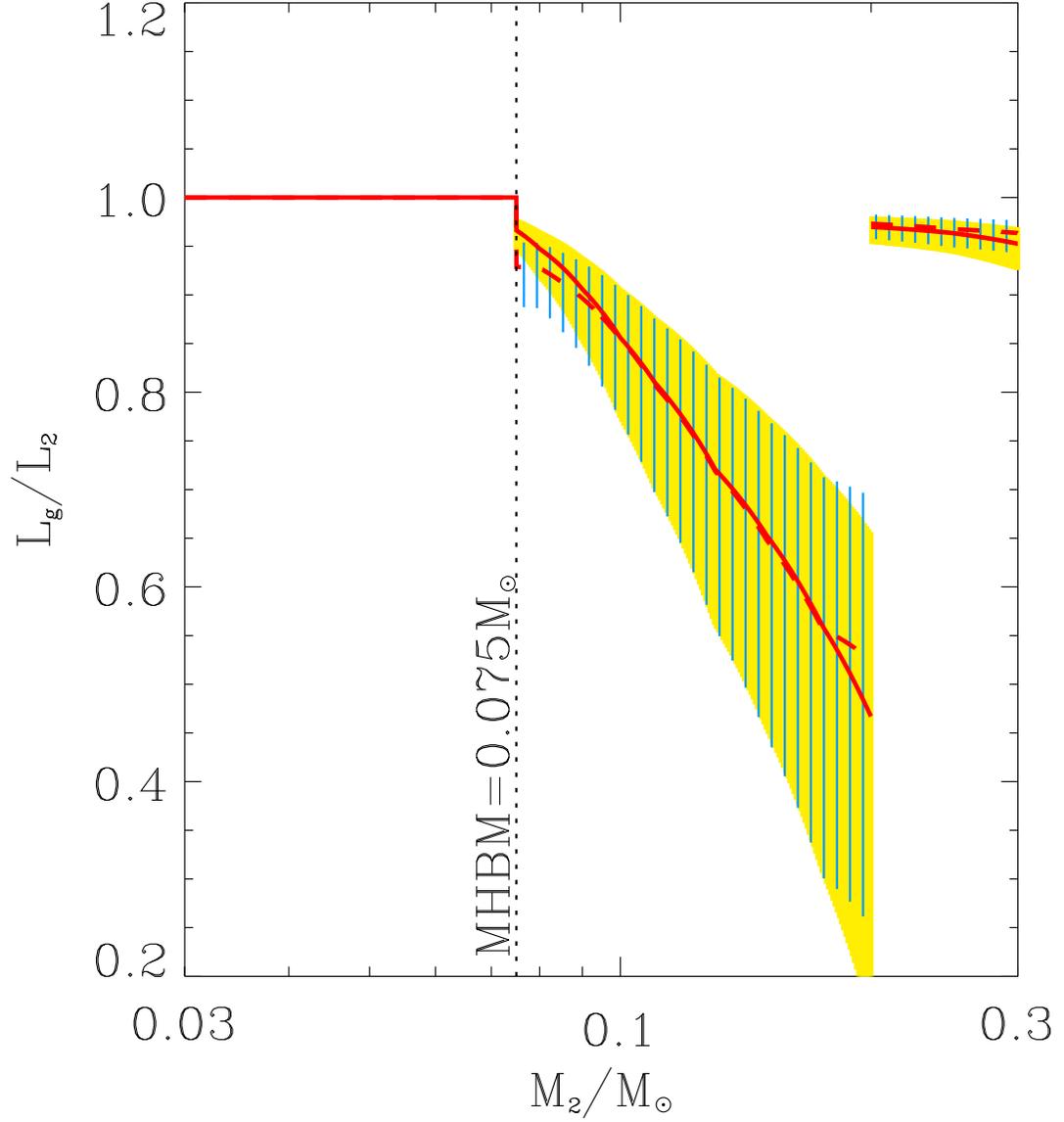}
 \caption{
Semi-empirical ratio of the gravothermal luminosity $\Lg$ to the
surface luminosity $L_2$ as a function of the donor mass. The dashed
and solid lines correspond to the B1 and B5 sequences, respectively,
with the shaded and hatched regions denoting the uncertainties. For
donors below the MHBM, all radiated energy comes from gravothermal
collapse ($\Lg/L_2=1$).
 }\label{pic_m_lg}
\end{figure}

Figure \ref{pic_m_lg} plots ${\Lg}/{L_2}$ versus $M_2$ for the B1
and B5 sequences as dashed and solid lines, respectively, with the
uncertainties denoted by the shaded and hatched regions. Figures
\ref{pic_m_al} and \ref{pic_m_lg} together imply that at any given
mass the contribution of the gravothermal luminosity increases with
the bloating factor. If $R_2$ increases by $20\%$, more than $80\%$
of the donor luminosity is generated by the release of thermal and
gravitational energies.

%fig4

\section{Mass-loss Rate}\label{sec_dM}

\par
Finding the theoretical mass-loss rates of CV donors is a daunting
task since it involves many complicated processes including thermal
and adiabatic relaxations, nuclear evolution, effects of tidal and
rotational distortions, orbital angular momentum loss, etc., all of
which may affect the internal structure significantly. Although the
stationary  model of \citet{c_Ritter_1988} has been used in various
theoretical models of CV evolution, it relies on the parameters
describing the donor atmosphere in the vicinity of the inner
Lagrangian point, which are poorly known. By relating the mass-loss
rate with angular momentum loss, on the other hand, the Ritter model
allows to check whether a proposed angular momentum loss mechanism
is indeed responsible for CV evolution. As mentioned in
\S\ref{sec:intro}, the numerical calculations that incorporate the
stellar evolution and the mass-loss model of \citet{c_Ritter_1988}
produced MRRs that differ from the empirical MRR.

\subsection{Semi-empirical Mass-loss Rate}\label{ssec_dM}

\par Instead of employing the Ritter model to calculate
the mass-loss rate, we seek the $\dM$ that is required to reproduce
the empirical MRR.  We make use of the well-known relation between
the effective mass-radius exponent $\xE$ of the donor and the
mass-loss timescale:
\begin{equation}\label{eq_zE_def}
 \xE=\xS+\tau\frac{7}{3}\frac{\Lg}{L_2},
\end{equation}
\citep{c_Ritter_1996}, where $\xS\approx -1/3$ is the adiabatic
mass-radius exponent for fully convective stars. Equation
(\ref{eq_zE_def}) dictates how the MRR should behave in response to
the mass-loss and thermal relaxation processes. When the mass
transfer just commences, $\Lg\approx0$ and the donor response is
adiabatic, leading to $\xE\approx\xS$. Since the ensuing deviation
from the adiabatic evolution depends on how the donor reacts
thermally on the mass loss, $\xE$ increases linearly with $\tau$
\citep{c_Ritter_1996}. Combining equations (\ref{eq_mean_den_law}),
(\ref{eq_st_bl}) and (\ref{eq_zE_def}) with the definitions of the
Kelvin-Helmholtz timescale ($\tKH\equiv GM_2^2/(R_2L_2$)) and
mass-loss timescale ($\tM\equiv M_2/\mid\dM\mid $), one obtains
\begin{equation}\label{eq_dm_def}
 -\dM=\frac{1}{(\xE+1/3)}\frac{56}{243}\frac{\sigma}{\pi}\frac{\Lg}{L_2}T_2^4 P^2.
\end{equation}
When $\xE=1/3$ corresponding to the period bounce, equation
(\ref{eq_dm_def}) recovers equation (34) of \citet{c_Rappaport_1982}
for the mass-loss rate at the minimum period.

\par
\begin{figure}
 \epsscale{0.9}
 \plotone{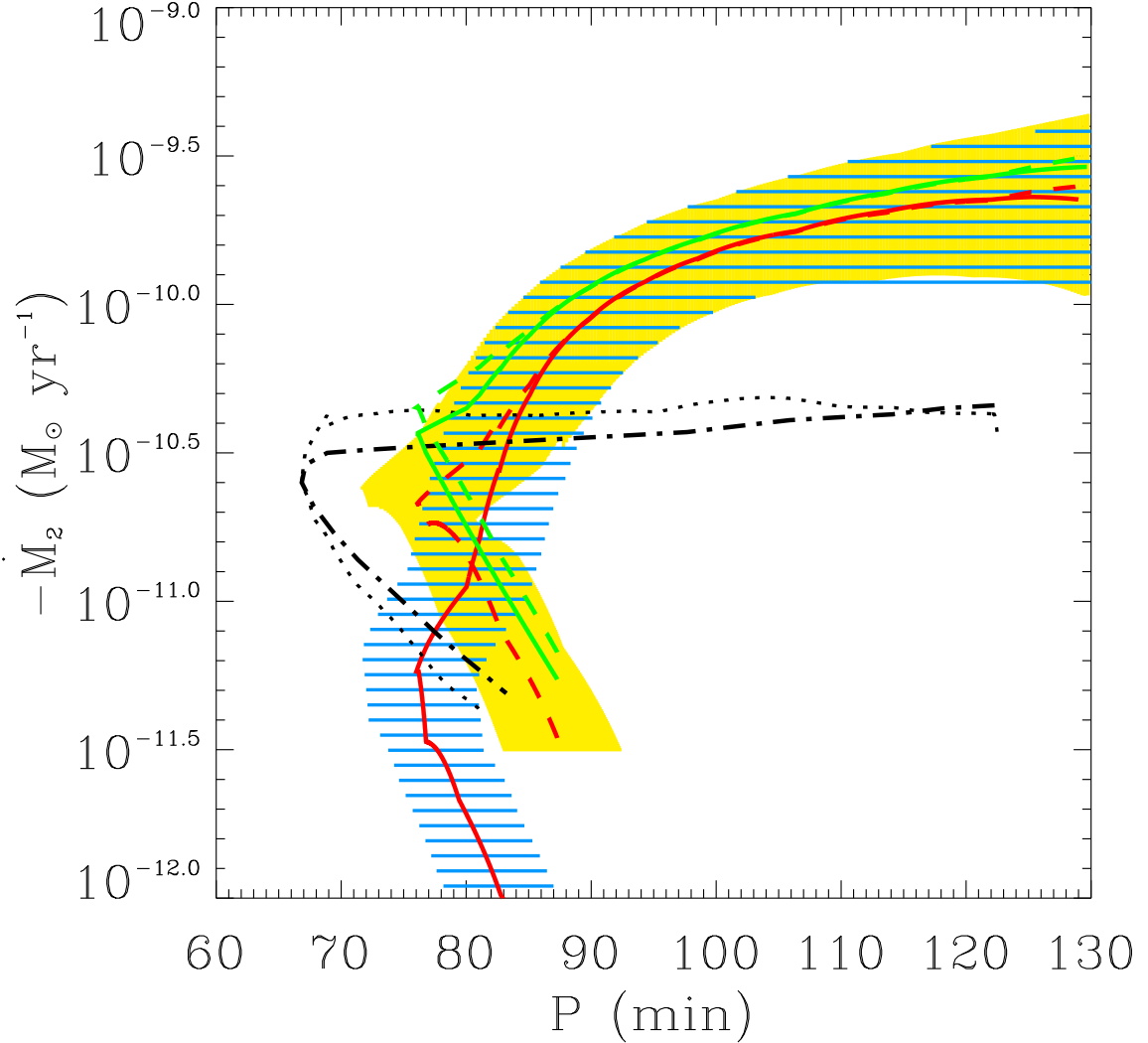}
 \caption{
Dependence of the mass-loss rate $\dM$ of a donor on the orbital
period $P$. The dashed and solid lines in red give $\dM$ for the B1
and B5 sequences, respectively, with the dashed and hatched regions
denoting the uncertainties. The dot-dashed line represents the
numerical result of \citet{c_Kolb_1999}, while the dotted line plots
the mass-loss rate obtained from equation (\ref{eq_dm_def}) with the
numerical MRR of \citet{c_Kolb_1999} for evolving donors. The dashed
and solid lines in green denote $\dMg$, for the B1 and B5 sequences,
respectively, from the stationary mass-transfer model of
\citet{c_Ritter_1988} when the angular momentum loss is dominated by
gravitational wave emission.
 }\label{pic_dM_P}
\end{figure}

In Figure \ref{pic_dM_P} we plot the semi-empirical $P$--$\dM$
relations calculated from (\ref{eq_dm_def}) for the 1-Gyr and 5-Gyr
isochrones as dashed and solid lines in red color, respectively. The
corresponding uncertainties are given by the shaded and hatched
regions. Columns (7) and (12) of Table \ref{ta_seq} list the values
of $\dM$. The semi-empirical mass-loss rate is essentially
non-stationary. For $129$ min $>P>90$ min, $\dM$ is in the range of
$\sim 10^{-9.5}$ -- $10^{-10}\dMd$ and decreases relatively slowly
as $\dM\propto P^3$. Below 90 minutes, $\dM$ sharply declines
towards $\Pmin$ as a result of a strong dependence of $\dM$ upon
$T_2$, emulating the $P$--$T_2$ relation shown in Figure
\ref{pic_Te_P}. Since the accretion rate decreases with $\dM$, this
rapid decline of $\dM$ is consistent with the observed low accretion
activities of CVs within the period spike identified in the SDSS
samples \citep{c_Gansike_2009}. Note that the mass-loss rates for
the B1 and B5 sequence are almost similar to each other for masses
above the MHBM, suggesting that these are perhaps permanent values.
In the brown dwarf regime, on the other hand, the ambiguity in the
adopted stellar age leads to a large difference in the mass-loss
rate, suggesting that it is necessary to use the appropriate
isochrones for an accurate estimation of $\dM$ for donors with
$M_2<\MHBM$.

\par For comparison, Figure  \ref{pic_dM_P} also plots as a dot-dashed
line the $P$--$\dM$ relation from the numerical models of
\citet{c_Kolb_1999}, which shows a pronounced ``boomerang'' shape
such that $\dM$ is nearly stationary above $\Pt$ with $\dM\propto
P^{0.5}$. It also has $\dM$ significantly smaller for $M_2>\MHBM$
than our semi-empirical results. To see what causes the difference
between our semi-empirical $P$--$\dM$ relation and the numerical
result of \citet{c_Kolb_1999}, we calculate $\dM$ from equation
(\ref{eq_dm_def}) by using the MRR of \citet{c_Kolb_1999} together
with the 5-Gyr isochrone. The resulting $P$--$\dM$ relation is
plotted in Figure \ref{pic_dM_P} as a dotted line, which is overall
in good agreement with the numerical results of \citet{c_Kolb_1999}.
This suggests that smaller radii of donors in their work are a
primary reason for the discrepancies between the results of the
current work and \citet{c_Kolb_1999}. This also proves to some
extent that our approach provides a reasonable way to estimate the
mass-loss rate.

%fig5

From the semi-empirical $\dM$, we calculate $\tau$ for the B1 and B5
sequences, which are listed in Columns (8) and (13) of Table
\ref{ta_seq}, respectively. Note that donors in CVs below $\Pgam$
are out of thermal equilibrium with $\tau\lesssim1$, which is the
main cause of the donor expansion. Equations (\ref{eq_Lg_def}) and
(\ref{eq_dm_def}) imply that for given $M_2$ and $\xE$, the
mass-loss rate increases more than $\alpha^3$ due mainly to changes
in the orbital period and gravothermal luminosity. The resulting
semi-empirical $P$--$\dM$ relation still has a boomerang shape,
which is much less dramatic than that of \citet{c_Kolb_1999} due to
the donor bloating and the associated thermal relaxation.
Nevertheless, it is very difficult to detect the boomerang shape
observationally because of the sharp decline of $T_2$ and $\dM$
below 90 min.

\par
\begin{figure}
 \plotone{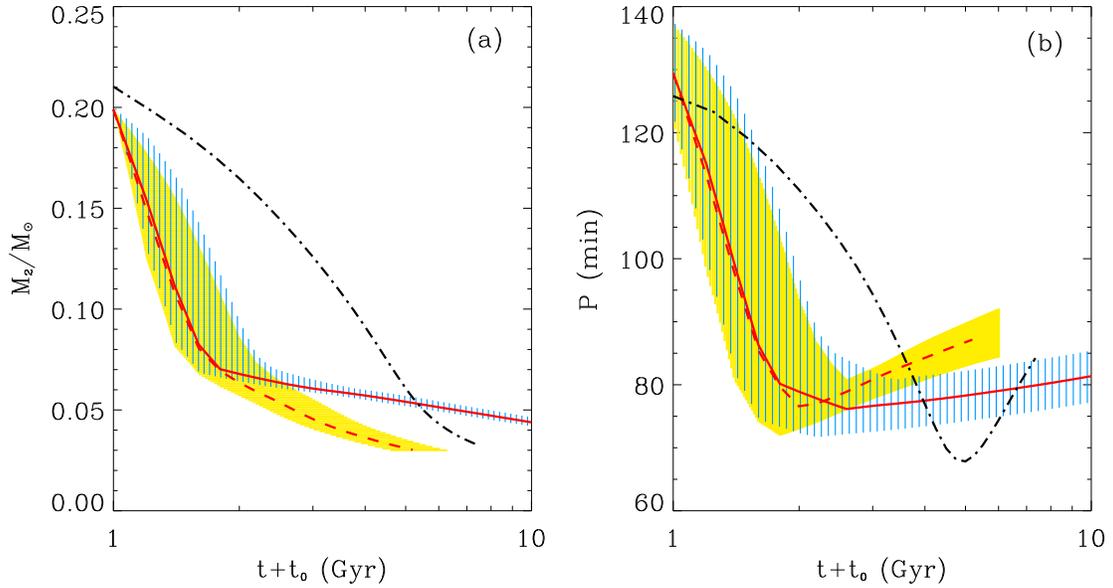}
 \caption{
Temporal evolution of (\emph{a}) the donor mass $M_2$ and (\emph{b})
the orbital period $P$ due to mass loss. Note that $t$ denotes the
time measured from the lower bound $\Pgam$ of the period gap where
the donor has an age of $t_0=1$ Gyr. The dashed and solid lines are
for the B1 and B5 sequences, respectively, with the shaded and
hatched regions denoting the uncertaities. The dot-dashed lines show
the results from the mass-loss rate of \citet{c_Kolb_1999}.
 }\label{pic_t_m}
\end{figure}

Once the $P$--$M_2$ and $P$--$\dM$ relations are found, it is
straightforward to calculate the temporal change of the donor mass
by solving $dM_2/dt = \dM$. We assume that donors are 1 Gyr old and
have a fixed initial mass of $0.2\Ms$ when they emerge from the
period gap.  Since CVs with $P < 2$ h are predominantly DNe
\citep{c_Ritter_2003}, we further assume that the masses of WDs are
constant at $M_1=0.75\Ms$ throughout CV evolution. Figure
\ref{pic_t_m} plots as dashed and solid lines the evolution of donor
mass and orbital period calculated from the semi-empirical $\dM$ for
the B1 and B5 sequences, respectively, as functions of time $t$
elapsed since $\Pgam$ (i.e., $t_0=1$ Gyr).  The dot-dashed lines
show $M_2(t)$ and $P(t)$ from the mass-loss rate of
\citet{c_Kolb_1999} for which $M_1=0.6\Ms$ and $M_2=0.21\Ms$
initially at $\Pgam$. The changes in the donor mass and orbital
period in the numerical model of \citet{c_Kolb_1999} are in general
slower because of a lower mass-loss rate. Our semi-empirical results
indicate that $M_2$ decreases from $0.2\Ms$ to $\MHBM$ in $\sim1$
Gyr and that the minimum period is reached in $\sim1$--$2$ Gyr after
$\Pgam$, without no essential dependence on the adopted stellar age.
On the other hand, the numerical model of \citet{c_Kolb_1999}
predicts that the same decrement of $M_2$ occurs in $\sim3.5$ Gyr,
while it takes the donors $\sim4$ Gyr to reach $\Pmin$.  Although
uncertainties in the stellar ages make it difficult to follow
$M_2(t)$ and $P(t)$ accurately below the MHBM, the close agreement
between the results based on the B1 and B5 sequences suggests that
the ages of CV secondaries with $\MHBM<M_2<0.2\Ms$ are less than 2
Gyr.

%fig6

\subsection{Consistency check}\label{ssec_dMg}

\par
Mass transfer in CVs below the period gap is considered to be driven
by gravitational wave radiation \citep{c_Pacz_1981,
c_Patterson_1984, c_Kolb_1999, c_Howell_2001}. While equation
(\ref{eq_dm_def}) is useful to obtain $\dM$ from the empirical MRR
independent of specifics of mass transfer, it does not provide any
clue regarding what drives mass transfer in CVs.  In this
subsection, we make use of the stationary mass-transfer model of
\citet{c_Ritter_1988} and check whether the semi-empirical $\dM$
calculated in \S\ref{ssec_dM} is consistent with the angular
momentum loss via emission of gravitational waves.

\citet{c_Ritter_1988} presented a simple analytical model in which
the mass transfer from a donor is treated as a stationary
isothermal, subsonic flow of gas through the inner Lagrangian point.
Taking allowance for the finite scale height of the donor
photosphere and considering the temporal changes in the volume radii
of the donor and the Roche lobe, \citet{c_Ritter_1988} showed that
the stationary mass-transfer rate is given by
\begin{equation}\label{eq_dm_Ritter}
 -\dMg=\frac{M_2}{(\xS-\xR)}
 \biggl[\frac{7}{3}\frac{\Lg}{L_2}\frac{1}{\tKH}-\frac{2}{\tJ}\biggr],
\end{equation}
where $\tJ$ is the angular momentum loss timescale, $\xR$ is the
mass-radius exponent of the critical Roche lobe. Note that we
neglect the nuclear-timescale term in equation (\ref{eq_dm_Ritter})
since it is much longer than the other timescales involved. In the
case of DNe for which the mass transfer is non-conservative, $\xR$
is given by
\begin{equation}
 \xR=-\frac{5}{3}+\biggl(\frac{2}{3}+2q\biggr)\frac{q}{1+q},
\end{equation}
\citep{c_Soberman_1997}. If the gravitational wave emission is the
dominant mechanism for the shrinkage of the orbital radius, $\tJ$
can be set equal to $\tG$ defined as
\begin{equation}\label{eq_tg}
 \tG=-\frac{5}{32}\frac{c^5}{G^{5/3}}
   \biggl(\frac{P}{2\pi}\biggr)^{8/3}
   \frac{(M_1+M_2)^{1/3}}{M_1M_2},
\end{equation}
\citep{c_Landau_1975}.

\par
We use the empirical MRR of \citet{c_Knigge_2006} and solve equation
(\ref{eq_dm_Ritter}) with $\tJ=\tG$ for $\dMg$. The resulting
$P$--$\dMg$ relations are plotted in Figure \ref{pic_dM_P} as dashed
and solid lines in green color for the B1 and B5 sequences.
respectively. The adopted stellar ages make a small difference in
$\dMg$, since the first term in the square brackets of equation
(\ref{eq_dm_Ritter}) is insensitive to the isochrones when $M_2>$
MHBM, while it becomes much smaller than the second term when $M_2<$
MHBM. Figure \ref{pic_dM_P} shows that $\dMg$ agrees, within
errorbars, with $\dM$ for both B1 and B5 sequences before reaching
the minimum period. This verifies that our semi-empirical mass-loss
rate is consistent with the angular momentum loss due to emission of
gravitational waves, at least for pre-bouncers. Beyond $\Pmin$,
$\dM$ for the B1 sequence appears to match $\dMg$ better than the B5
sequence, suggesting that the CV ages are presumably less than 5
Gyr.

\section{Discussion}\label{sec_Dis}

\par
\begin{figure}
 \plotone{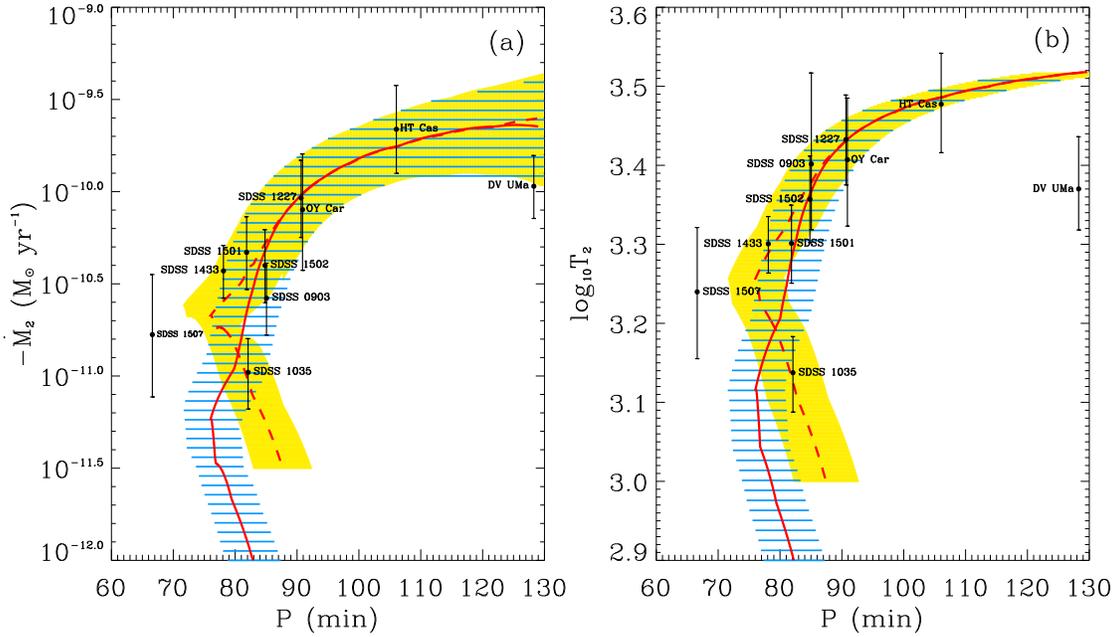}
 \caption{
Comparison between our semi-analytic results (\emph{lines}) and the
10 observed eclipsing CVs (\emph{filled circles}) listed in Table
\ref{ta_ecl} for (\emph{a}) the mass-loss rate and (\emph{b}) the
effective temperature. The dashed and solid lines are for the B1 and
B5 sequences, respectively, with the shaded and hatched regions
denoting the uncertaities. In (\emph{b}), $T_2$ is estimated from
equation (\ref{eq_T1_T2}) assuming that mass is conserved (i.e.,
$-\dM=\adMwd$).
 }\label{pic_dm_te_ecl}
\end{figure}

Since the mass-loss rate cannot be observed directly, we are only
able to compare our semi-empirical results with the accretion rates
estimated by means of other indicators. For this purpose, we select
10 short-period, eclipsing CVs from \citet{c_Patterson_2005},
\citet{c_Littlefair_2008}, and \citet{c_Townsley_2009}. Table
\ref{ta_ecl} represents a compilation of our sample CVs, with the
parameters in Columns (2) to (7) taken from the referred papers. To
our knowledge, these are the only eclipsers with $P$ below 130 min,
for which the reliable data for the binary parameters such as $P$,
$M_1$, $M_2$, and $T_1$ are all available. To infer the
mass-accretion rates for the sample CVs, we use
\begin{equation}\label{eq_Town_2009}
 T_1=1.7\times10^{4} K
 \biggl(\frac{\adMwd}{10^{-10}\dMd}\biggr)^{1/4}
 \left(\frac{M_1}{0.9\Ms}\right),
\end{equation}
where $T_1$ and $M_1$ are the effective temperature and mass of a WD
and $\adMwd$ is the time-averaged accretion rate, taken from
\citet{c_Townsley_2009}. Since the gravitational energy released by
the accreted material onto the surface of a WD is deposited at
shallow depths and soon radiated away, the quiescent surface
luminosity should be proportional to the mass-accretion rate,
leading to $T_1\propto \adMwd^{1/4}$ (e.g.,
\citealt{c_Townsley_2003,c_Townsley_2004}). Filled circles with
errorbars in Figure \ref{pic_dm_te_ecl}\emph{a} plot $\adMwd$ for
the sample CVs calculated from equation (\ref{eq_Town_2009}); these
are also listed in Column (8) of Table \ref{ta_ecl}. It can be seen
that $\adMwd$ is broadly in line with our semi-empirical $\dM$ for
both B1 and B5 sequence. This suggests that our semi-empirical
$P$--$\dM$ relation is realistic, provided that $-\dM=\adMwd$ in
these CVs during accretion disk quiescence. \citet{c_Townsley_2009}
noted that $\adMwd$ significantly declines towards $\Pmin$ for
$P<84$ min. Since the accretion rate decreases with decreasing
$\dM$, this downward tendency of $\adMwd$ with $P$ is in agreement
with the behavior of the semi-empirical $\dM$.

%fig7

\par
Donors in short-period CVs are very faint objects, making it
difficult to measure their effective temperatures $T_2$ directly
from observations, especially in the brown dwarfs regime: in our
sample, only two systems (HT Cas and DV UMa) have measured $T_2$
available in the \citet{c_Ritter_2003} catalogue. One way to
estimate $T_2$ is to use the strong dependence of $\dM$ on $T_2$.
Assuming that $-\dM=\adMwd$ during the quiescence phase of DNe, we
combine equations (\ref{eq_dm_def}) and (\ref{eq_Town_2009}) to
derive
\begin{equation}\label{eq_T1_T2}
 \frac{T_2}{T_1}\approx11.594\frac{(\xE+1/3)^{1/4}}{P^{1/2}}
 \biggl(\frac{\Lg}{L_2}\biggr)^{-1/4}
 \biggl(\frac{M_1}{0.9\Ms}\biggr)^{-1}.
\end{equation}
Since donors with mass below the MHBM have $\Lg/L_2=1$,
${T_2}/{T_1}$ depends only on the orbital period and the effective
mass-radius exponent. For masses above the MHBM, on the other hand,
an accurate estimation of $T_2$ requires to consider the effect of
$(\Lg/L_2)^{-1/4}$ term, which can not be determined
observationally.

\par Using the method described in \S\ref{ssec_Lg}, we calculate
$\Lg/L_2$ for the sample CVs listed in Table \ref{ta_ecl}. For this,
we take $\xE=0.21\pm0.1$ for systems with $M_2<0.063\Ms$ and
$\xE=0.64\pm0.02$ for $M_2> 0.063\Ms$ in accordance with the adopted
empirical MRR. The bloating factor of each system is calculated
relative to the 1-Gyr isochrone. We then calculate $T_2$ from
equation (\ref{eq_T1_T2}) and plot the results in Figure
\ref{pic_dm_te_ecl}\emph{b} as filled circles with errorbars; these
are also tabulated in Column (9) in Table \ref{ta_ecl}. It is
remarkable that the estimated effective temperatures, except for
SDSS 1507 and DV UMa, are very close to the semi-empirical
$P$--$T_2$ relation, suggesting that equation (\ref{eq_T1_T2}) is in
fact a good way to estimate $T_2$ when it cannot be measured
observationally. A small orbital period of SDSS 1507 implies that
this system is most likely formed directly from a WD/brown dwarf
pair \citep{c_Littlefair_2007} or a member of the old halo
\citep{c_Patterson_2008_k}, rather than a conventional DN. In the
case of DV UMa, $T_2\approx 2400$ K calculated from equation
(\ref{eq_T1_T2}) is $\sim 30\%$ smaller than the observed effective
temperature  $T_2\approx3200\pm100$ K, corresponding to a spectral
type of M4.5$\pm0.5$, from the \citet{c_Ritter_2003} catalogue. Such
a big difference in $T_2$ cannot be due to ambiguity in
$(\xE+1/3)^{1/4}$ or $(\Lg/L_2)^{-1/4}$ terms since both are close
to unity for this object. \citet{c_Townsley_2003} noted that the
best-constrained quantity from a measured $T_1$ is the accretion
rate per unit area, $\langle\dot
m_1\rangle=\langle\dot{M_1}\rangle/4\pi R_1^2$, while the subsequent
estimation of $\adMwd$ from $\langle\dot m_1\rangle$ requires a
certain MRR of WDs, which is poorly known for CVs. The WD mass in DV
UMa is highest  among the sample, and it is highly probable that the
MRR for massive WDs differs from $R_1\ \propto M_1^{-1.8}$ adopted
by \citet{c_Townsley_2009}.

\par
Based on the donor masses, \citet{c_Littlefair_2008} claimed that
SDSS 1035, SDSS 1501, SDSS 1507, and SDSS 1433 possess post-period
minimum CVs. Although period bouncers should have low $M_2$, the
precise mass of a particular secondary does not signify its
evolutionary history or status \citep{c_Patterson_2008}. The
evolutionary status of such objects can be clarified more clearly by
their positions, for instance, on the $P$--$T_2$ or $P$--$\dM$
diagram where the boomerang effect can be seen. Figure
\ref{pic_dm_te_ecl} suggests that SDSS 1501 and SDSS 1433 among the
sample of \citet{c_Littlefair_2008} have not reached $\Pmin$ yet.
This result does not depend on the stellar age, since these have
$-\dM> 10^{-10.5}\dMd$ for which the semi-empirical mass-loss rate
is insensitive to the adopted isochrone. While the evolutionary
status of SDSS 1035 is unclear, the fact that the $P$--$\dM$
dependence for the B1 sequence reconciles well with that due to
gravitational wave emission indicates that it is likely a
post-bounce CV.

\par
\citet{c_Kolb_1999} argued that the tidal and the rotational
distortions do not change the value of $\Pt$ appreciably because the
internal structure of CV secondaries is only weakly affected by such
perturbations. To show this, they compared the effective
temperatures of the distorted and undistorted stars and found almost
no change in $T_2$. This result is totally expected since
short-period CVs are fully convective stars. As we showed in
\S\ref{ssec_Te}, the Hayashi theory gives that the reaction of $T_2$
on the bloating of a donor is small for isolated, fully convective
stars. We also show in Appendix \ref{ap_Hayashi_mod} that the
influence of an isentropic expansion on $T_2$ caused by tidal and
rotational distortions can be neglected in most practical cases. All
of these suggest that the effective temperature does not reliably
measure the amount of the stellar bloating, and consequently can not
be used as an indicator of the degree of tidal and rotational
distortions.

\section{Summary}\label{sec_Con}

\par
The mass-loss rate of donors is one of the most important parameters
that govern the evolution of short-period CVs. Since the mass-loss
rate is not directly observable, most previous work employed either
purely numerical methods or semi-empirical approaches that use the
observed properties of CV accretors rather than donors. Yet, the
confirmation of the resulting mass-loss rates has to be made. In
this work, we use the empirical MRR of short-period CVs constructed
by \citet{c_Knigge_2006} as an input parameter and derive the
semi-empirical relationship between the orbital period and mass-loss
rate. We assume that the bloating of CV donors is caused mainly by
the mass loss and associated thermal relaxation processes. We also
calculate the responses of the effective temperature and
luminosities on the bloating of the CV secondaries. We discuss our
semi-empirical mass-loss rate in comparison with those from the
previous numerical and semi-empirical studies, and in terms of
estimating the effective temperature of the donors. Our main results
are summarized as follows.

\par
1. The semi-empirical $P$--$\dM$ dependence shows that the mass-loss
rate is essentially non-stationary. For $P>90$ min, $\dM$ is in the
range of $10^{-9.5}$--$10^{-10}\dMd$ and decreases with decreasing
$P$ relatively slowly as $\dM\propto P^3$. Below 90 minutes, on the
other hand, $\dM$ sharply declines towards $\Pmin$ as a result of a
strong dependence of $\dM$ upon $T_2$, emulating the $P$--$T_2$
relation. The $P$-dependence and amplitudes of the semi-empirical
$\dM$ are significantly different from those predicted by the purely
numerical models of \citet{c_Kolb_1999}. The main reason for the
discrepancies is small donor radii in the latter. Donors in
short-period CVs are out of thermal equilibrium with
$\tau\lesssim1$. Thermal relaxation processes change the slope of
the $P$--$\dM$ and $P$--$T_2$ diagrams significantly above $\Pmin$,
making the boomerang shape less pronounced. Due to the sharp decline
of $T_2$ and $\dM$ below 90 min, the observational detection of the
boomerang shape on the $P$--$\dM$ and $P$--$T_2$ is highly unlikely.

\par
2. The semi-empirical $P$--$\dM$ relation on the basis of the 1-Gyr
isochrone is consistent with that estimated from the mass-transfer
model of \citet{c_Ritter_1988} with the angular momentum carried
dominantly by gravitational waves. Our semi-empirical $P$--$\dM$
relation predicts that after emerging from the period gap, donors
lose mass from $0.2\Ms$ to $\MHBM$ in $\sim1$ Gyr, and reach the
minimum period in $\sim1$--2 Gyrs, suggesting that the ages of CV
secondaries with $\MHBM<M<0.2\Ms$ are less than 2 Gyrs.

\par
3. For 10 selected eclipsing CVs (listed in Table \ref{ta_ecl}) for
which the reliable observational data are available, the accretion
rates inferred from the observed effective temperatures of WDs are
in good agreement with our semi-analytic mass-loss rates, indicating
that our semi-empirical $P$--$\dM$ relation is realistic. This also
suggests that the effective temperature of a CV secondary can be
estimated from the effective temperature of its WD primary through
equation (\ref{eq_T1_T2}). When analyzed on the $P$--$\dM$ or
$P$--$T_2$ plane, two CVs (SDSS 1501 and SDSS 1433) that were
previously considered as post-bouncers by \citet{c_Littlefair_2008}
are most likely to be systems before reaching the minimum period.

\acknowledgments
The authors are grateful to the referee, Dr.\ J.\ Patterson,
for a thoughtful report.
This work was supported by the National Research
Foundation of Korea (NRF) grant funded by the  Korean government
(MEST), No. 2009-0063616.

\appendix{}
\section{Isentropic Expansion/Contraction of Polytropic Donors}\label{ap_isentropic}

Even without considering the effect of mass loss,
\citet{c_SWTK_2009} showed that the tidal and rotational distortions
alone make donors in CVs larger by about $\sim 4$--$8\%$ for
$n=1.5$ polytropes and $\sim2$--$3\%$ for $n=3.5$ polytropes,
depending on the mass ratio. In this Appendix, we evaluate the
enlargement factor for polytropes with arbitrary $0<n<5$.

Consider a polytropic star with index $n$, mass $M_2$, and pressure
constant $K$. Its radius is $R_{2,0}$ in isolation. The MRR of the
unperturbed polytropes is given by
\begin{equation}\label{eq_poly_MRR} K = N_{n,0} G
M_2^{(n-1)/n} R_{2,0}^{(3-n)/n},
\end{equation}
where $N_{n,0}$ is the dimensionless pressure constant that varies
weakly with $n$ \citep{cha33a}. When it is situated in a CV, the
presence of tidal and rotational perturbations causes it to expand
or contract, depending on $n$, by adjusting the internal structure
to restore hydrostatic equilibrium. Let the new equilibrium be
achieved at the volume radius $R_2$ with the corresponding
dimensionless pressure constant $N_n$. Since the global internal
adjustment occurs in the dynamical timescale, shorter than the
thermal timescale and the mass-loss time scale, one may assume that
the specific entropy of the gas inside the donor does not change
much under the influence of the distortional perturbations.

\begin{figure}
 \epsscale{0.9}
 \plotone{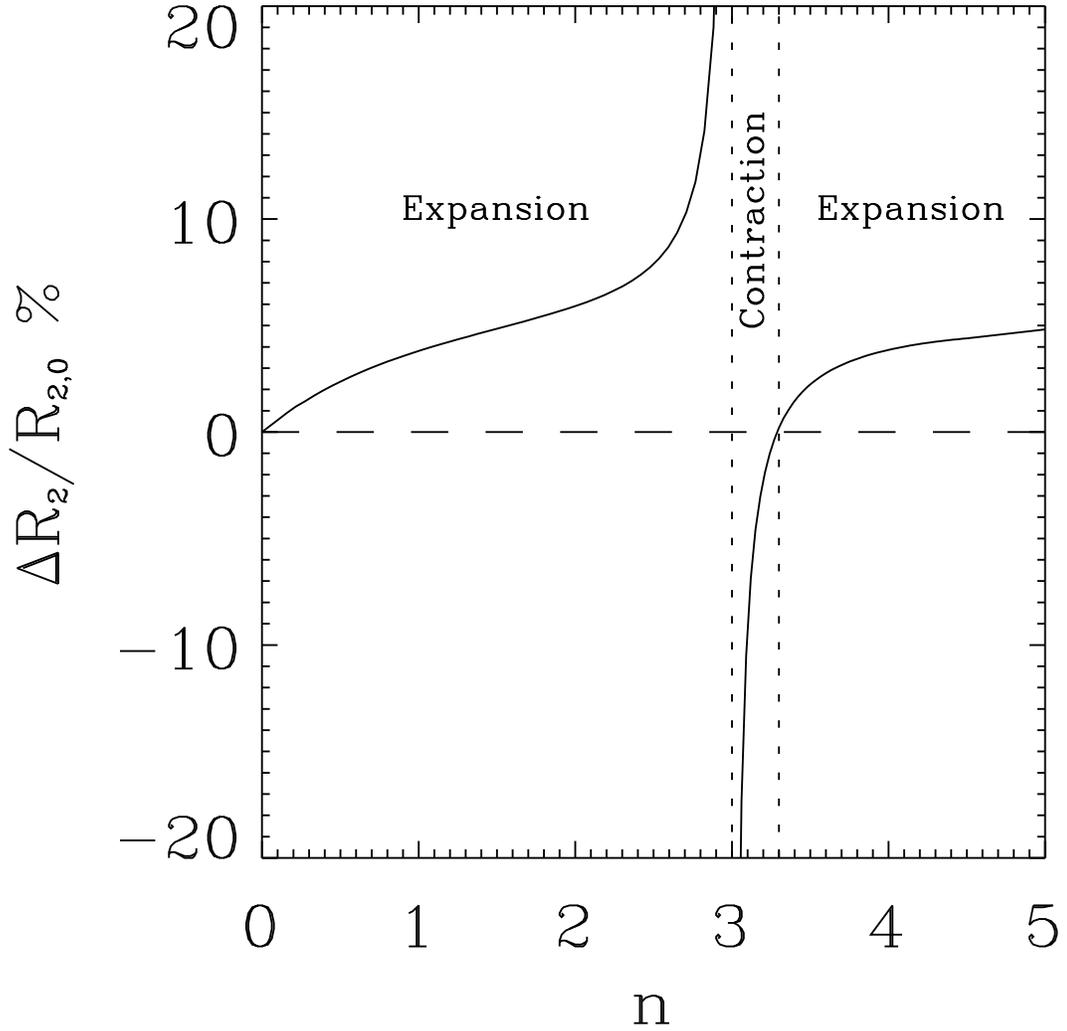}
 \caption{
Changes in the volume radius of a polytrope due to the isentropic
expansion under tidal and rotational perturbations against the
polytropic index $n$, when the mass ratio is $q=0.1$. A binary
component expands when $n<3$ or $n\simgt 3.3$,  while it contracts
when $3<n\simlt 3.3$. Note that $n=3$ is a special case where the
size of a polytrope cannot be constrained by its mass.
 }\label{pic_npl}
\end{figure}

For isentropic responses (i.e., $K$ is unchanged), the enlargement
factor is given simply by
\begin{equation}
 \Aise \equiv\frac{R_2}{R_{2,0}} =
 \biggl(\frac{N_n}{N_{n,0}}\biggr)^{n/(n-3)}.
\end{equation}
We use a self-consistent field method outlined in
\citet{c_SWTK_2009} to construct detailed internal structure of
polytropes in hydrostatic equilibrium.  We then calculate $N_{n,0}$
and $N_n$ for polytropes in critical (i.e., Roche-lobe filling)
configurations by varying $n$ and the mass ratio $q=M_2/M_1$. Figure
\ref{pic_npl} plots the resulting relative changes, $\Delta
R_2/R_{2,0} = \Aise-1$, in the stellar size as a function of $n$ for
$q=0.1$.  The behavior of $\Aise$ with $M_2$ (or with the mass ratio
$q$ for $M_1=0.75\Ms$) for $n=3/2$ polytropes is plotted in Figure
\ref{pic_m_al} as a dotted line. Figure \ref{pic_npl} shows that
$\Delta R_2>0$ for $n<3$ or $n\simgt 3.3$, indicating that the
perturbations make the donors bigger; the increase in the equatorial
radius is larger than the decrease in the polar radius.  On the
other hand, critical polytropes with $3<n \simlt 3.3$ have $\Delta
R_2 <0$, since the perturbations decrease the polar radius much more
than increasing the equatorial radius.

Notice a discontinuity at $n=3$ for which one cannot determine the
stellar enlargement factor. For these polytropes, the central
density is proportional to $R_2^{-3}$ and the mass becomes
independent of $R_2$, making it impossible to constrain the mass
from its radius (e.g., \citealt{cha33a}). $\Delta R_2$ diverges to
either positive or negative infinity as $n\rightarrow 3$, indicating
that a small change of $n$ near 3 leads to a big difference in
$\Delta R_2$. This suggests that one should be cautious when
treating $n=3$ polytropes numerically. Using SPH simulations,
\citet{c_Renvoize_2002} found that the critical configurations have
an enlargement factor $\sim10$--$12\%$ for $n=3$ polytropes when
$q=0.05$--$1.0$. This is presumably due to some (unavoidable)
numerical errors in the SPH simulations that somehow change their
$n=3$ polytropes to $n\sim 2.7$--$2.8$ ones ``effectively'',
resulting in $\Delta R_2/R_{2,0} = 0.10$--$0.12$.

%fig8

\section{Change of Effective Temperature due to Isentropic Expansion }\label{ap_Hayashi_mod}

\par
In \S\ref{ssec_Te}, we describe how the effective temperature $T_2$
of a fully-convective donor varies due to a homologous expansion
using the Hayashi theory. Here we study how $T_2$ responds to an
isentropic expansion. We assume that the pressure $P_i$ in the whole
stellar interiors below photosphere depends on temperature as
\begin{equation}\label{eq_P_int_def}
 P_i=\frac{\Re^{5/2}T^{5/2}}{\mu^{5/2}K^{3/2}},
\end{equation}
where $\mu$ is the mean molecular weight and $\Re$ is the gas constant.
The photosphere is determined at the radius where the optical depth is
$\tau_{\rm ph}=2/3$. If the radiative pressure is unimportant, the
photospheric pressure, $P_p$, is given by
\begin{equation}
 P_p(\tau_{\rm ph})\approx\frac{2}{3}\frac{\bg}{\kappa},
\end{equation}
where $\kappa$ is the opacity and $\bg=GM_2/R_2^2$ is the mean surface
gravity.  Using a simple opacity law of the form
$\kappa=\kappa_{0}P^a_{p} T^b$, we obtain
\begin{equation}\label{eq_P_ph_def}
 P_p(\tau_{\rm ph})\approx
 \biggl(\frac{2}{3}\frac{\bg}{\kappa_0}\frac{1}{T^{b}}\biggr)^{\frac{1}{a+1}}.
\end{equation}

\par Since we assume that the donor is in hydrostatic
equilibrium,  the photospheric pressure should equal the interior
pressure at the interface between the convective interiors and the
radiative photospheric layer.  Assuming $K=K_0$, equations
(\ref{eq_P_int_def}) and (\ref{eq_P_ph_def}) are combined to yield
\begin{equation}\label{eq_Te_cor}
 \frac{T_2}{T_{2,0}}=\Aise^{-\frac{4}{5a+2b+5}}.
\end{equation}
In contrast to the Hayashi theory, the effective temperature
\emph{decreases} due to the donor expansion. Nevertheless,
$T_2/T_{2,0}=0.99$ for $\Aise=1.05$, $a=0.5$, and $b=4.5$, typical
values for CV secondaries, so that the change in $T_2$ due to the
tidal and rotational distortions are still very small and can thus
be neglected in practice.

\begin{deluxetable}{cccccccccccccc}
 \tabletypesize{\scriptsize}
 \tablecaption {The semi-empirical donor sequence for CVs}
 \tablewidth{0pt}
 \tablehead {&&&\multicolumn{5}{c}{1-Gyr}&&\multicolumn{5}{c}{5-Gyr}\\
 \cline{4-8}\cline{10-14}\\
 $M_2$& $R_2$&$P$&$t+t_0$&$\log L_2$&$T_2$&$\log\dM$&$\tau$&&$t+t_0$&$\log L_2$&$T_2$&$\log\dM$&$\tau$ \\
 \\
 \colhead{(1)} & \colhead{(2)} & \colhead{(3)} & \colhead{(4)} & \colhead{(5)} &
 \colhead{(6)} & \colhead{(7)}  & \colhead{(8)} && \colhead{(9)} &
 \colhead{(10)} & \colhead{(11)}& \colhead{(12)}& \colhead{(13)}}
 \startdata
0.040&0.099& 82.769& 3.50&-4.625&1273&-11.10&0.24&&13.08&-5.513& 763&-11.99&0.24\\
0.044&0.101& 81.252& 3.10&-4.435&1407&-10.94&0.23&& 9.74&-5.317& 846&-11.83&0.23\\
0.048&0.103& 79.898& 2.80&-4.293&1513&-10.83&0.23&& 7.31&-5.167& 914&-11.71&0.23\\
0.053&0.105& 78.676& 2.55&-4.194&1588&-10.76&0.23&& 5.47&-5.011& 992&-11.58&0.23\\
0.057&0.107& 77.565& 2.33&-4.121&1645&-10.73&0.24&& 4.08&-4.880&1062&-11.49&0.24\\
0.063&0.108& 76.021& 2.05&-3.945&1804&-10.67&0.29&& 2.64&-4.507&1305&-11.23&0.29\\
0.065&0.111& 77.472& 1.95&-3.828&1905&-10.61&0.32&& 2.30&-4.316&1439&-11.09&0.32\\
0.069&0.116& 79.730& 1.80&-3.679&2035&-10.56&0.40&& 1.87&-4.108&1590&-10.99&0.40\\
0.075&0.121& 82.611& 1.70&-3.481&2226&-10.39&0.41&& 1.72&-3.649&2021&-10.55&0.41\\
0.078&0.124& 84.036& 1.65&-3.384&2327&-10.33&0.44&& 1.67&-3.461&2226&-10.39&0.43\\
0.082&0.128& 86.094& 1.59&-3.255&2466&-10.21&0.44&& 1.60&-3.282&2427&-10.23&0.44\\
0.086&0.133& 88.096& 1.56&-3.146&2584&-10.11&0.45&& 1.57&-3.156&2570&-10.12&0.44\\
0.090&0.137& 90.046& 1.53&-3.063&2671&-10.04&0.46&& 1.54&-3.064&2668&-10.04&0.45\\
0.095&0.141& 91.948& 1.50&-2.987&2751& -9.98&0.46&& 1.52&-2.986&2751& -9.98&0.46\\
0.099&0.144& 93.805& 1.47&-2.925&2812& -9.94&0.47&& 1.49&-2.924&2813& -9.93&0.47\\
0.103&0.148& 95.619& 1.44&-2.870&2865& -9.90&0.48&& 1.46&-2.869&2866& -9.89&0.48\\
0.107&0.152& 97.394& 1.41&-2.821&2908& -9.86&0.49&& 1.43&-2.820&2910& -9.86&0.49\\
0.111&0.156& 99.132& 1.38&-2.778&2946& -9.84&0.51&& 1.40&-2.777&2948& -9.83&0.50\\
0.116&0.160&100.835& 1.36&-2.736&2983& -9.81&0.52&& 1.38&-2.735&2984& -9.81&0.52\\
0.120&0.163&102.504& 1.34&-2.699&3012& -9.79&0.53&& 1.36&-2.698&3014& -9.79&0.53\\
0.124&0.167&104.142& 1.32&-2.666&3037& -9.77&0.54&& 1.34&-2.665&3038& -9.77&0.54\\
0.128&0.170&105.751& 1.30&-2.636&3057& -9.76&0.56&& 1.32&-2.636&3058& -9.76&0.56\\
0.132&0.174&107.331& 1.28&-2.605&3080& -9.74&0.57&& 1.30&-2.605&3081& -9.74&0.57\\
0.137&0.177&108.885& 1.26&-2.575&3104& -9.73&0.58&& 1.28&-2.574&3105& -9.72&0.58\\
0.141&0.181&110.413& 1.24&-2.547&3124& -9.71&0.59&& 1.26&-2.546&3125& -9.71&0.59\\
0.145&0.184&111.916& 1.22&-2.521&3141& -9.70&0.61&& 1.24&-2.520&3143& -9.70&0.60\\
0.149&0.188&113.396& 1.20&-2.497&3155& -9.69&0.62&& 1.22&-2.496&3157& -9.69&0.62\\
0.154&0.191&114.854& 1.18&-2.472&3174& -9.68&0.64&& 1.20&-2.471&3176& -9.68&0.63\\
0.158&0.194&116.291& 1.16&-2.447&3191& -9.67&0.65&& 1.18&-2.446&3193& -9.67&0.65\\
0.162&0.198&117.706& 1.15&-2.425&3206& -9.66&0.67&& 1.16&-2.424&3208& -9.66&0.66\\
0.166&0.201&119.103& 1.13&-2.403&3219& -9.66&0.68&& 1.15&-2.403&3220& -9.65&0.68\\
0.170&0.204&120.480& 1.11&-2.383&3230& -9.65&0.70&& 1.13&-2.383&3231& -9.65&0.70\\
0.175&0.207&121.839& 1.10&-2.365&3240& -9.65&0.72&& 1.11&-2.364&3241& -9.65&0.72\\
0.179&0.211&123.180& 1.08&-2.345&3252& -9.64&0.73&& 1.09&-2.344&3253& -9.64&0.74\\
0.183&0.214&124.505& 1.06&-2.326&3263& -9.63&0.74&& 1.07&-2.325&3264& -9.64&0.76\\
0.187&0.217&125.813& 1.05&-2.308&3272& -9.62&0.75&& 1.05&-2.307&3274& -9.64&0.78\\
0.191&0.220&127.105& 1.03&-2.291&3281& -9.61&0.75&& 1.03&-2.290&3283& -9.64&0.81\\
0.196&0.223&128.382& 1.01&-2.275&3289& -9.60&0.76&& 1.02&-2.274&3290& -9.64&0.84\\
0.200&0.226&129.563& 1.00&-2.260&3295& -9.60&0.78&& 1.00&-2.260&3296& -9.65&0.88\\
 \enddata
\label{ta_seq} \tablecomments{$M_2$ and $R_2$ are the mass and
radius of the donor in the solar units, respectively; $P$ is the
orbital period in minutes; $t$ is the time in Gyr elapsed since the
lower bound $\Pgam$ of the period gap, with $t_0=1$ Gyr being the CV
age at $\Pgam$;  $T_2$ is the effective temperature of the donor in
degrees Kelvin; $L_2$ is the luminosity of the donor in L$_\odot$;
$\dM$ is the mass-loss rate in units of $\dMd$; $\tau$ is the
dimensionless parameter defined in equation (\ref{eq_t_def}).}
\end{deluxetable}

\begin{deluxetable}{lrcccccccc}
 \tabletypesize{\scriptsize}
 \rotate
 \tablecaption {Data on Eclipsing Cataclysmic Variables ($P<130$ min) with estimated effective temperature of WD}
 \tablewidth{0pt}
 \tablehead {
 \colhead{System}&\colhead{$P$}  &\colhead{$M_1$}&
 \colhead{$q$}   &\colhead{$T_1$}&
 \colhead{$M_2$}&\colhead{$R_2$}&\colhead{$\dMwd$}&\colhead{$T_2$} & \colhead{Ref} \\
 \colhead{(1)} & \colhead{(2)} & \colhead{(3)} & \colhead{(4)} & \colhead{(5)} &
 \colhead{(6)} & \colhead{(7)} & \colhead{(8)} & \colhead{(9)} &
 \colhead{(10)}
}
 \startdata
  SDSS 1507&   66.6119&$0.910^{+0.070}_{-0.070}$&$0.0625^{+0.0004}_{-0.0004}$&$11000^{+ 1250}_{- 1250}$&$0.057^{+0.004}_{-0.004}$&$0.097^{+0.003}_{-0.003}$&$-10.78^{+  0.33}_{-  0.34}$&$ 1730^{+  360}_{-  310}$&3\\\\
 SDSS 1433&   78.1066&$0.868^{+0.007}_{-0.007}$&$0.0690^{+0.0030}_{-0.0030}$&$12800^{+  950}_{-  950}$&$0.060^{+0.003}_{-0.003}$&$0.109^{+0.003}_{-0.003}$&$-10.43^{+  0.14}_{-  0.15}$&$ 1990^{+  160}_{-  160}$&3\\\\
 SDSS 1501&   81.8513&$0.800^{+0.030}_{-0.030}$&$0.0670^{+0.0030}_{-0.0030}$&$12500^{+  950}_{-  950}$&$0.053^{+0.003}_{-0.003}$&$0.108^{+0.004}_{-0.004}$&$-10.33^{+  0.19}_{-  0.20}$&$ 2000^{+  240}_{-  220}$&3\\\\
 SDSS 1035&   82.0896&$0.940^{+0.010}_{-0.010}$&$0.0550^{+0.0020}_{-0.0020}$&$10100^{+ 1000}_{- 1000}$&$0.052^{+0.002}_{-0.002}$&$0.108^{+0.003}_{-0.003}$&$-10.98^{+  0.18}_{-  0.20}$&$ 1370^{+  150}_{-  150}$&3\\\\
 SDSS 1502&   84.8298&$0.820^{+0.030}_{-0.030}$&$0.1090^{+0.0030}_{-0.0030}$&$12300^{+  950}_{-  950}$&$0.090^{+0.004}_{-0.004}$&$0.131^{+0.003}_{-0.003}$&$-10.40^{+  0.19}_{-  0.20}$&$ 2260^{+  300}_{-  270}$&3\\\\
 SDSS 0903&   85.0659&$0.960^{+0.030}_{-0.030}$&$0.1170^{+0.0030}_{-0.0030}$&$13000^{+ 1050}_{- 1050}$&$0.112^{+0.004}_{-0.004}$&$0.141^{+0.003}_{-0.003}$&$-10.58^{+  0.19}_{-  0.20}$&$ 2470^{+  740}_{-  430}$&3\\\\
 SDSS 1227&   90.6610&$0.810^{+0.030}_{-0.030}$&$0.1180^{+0.0030}_{-0.0030}$&$15000^{+ 1250}_{- 1250}$&$0.096^{+0.004}_{-0.004}$&$0.140^{+0.003}_{-0.003}$&$-10.03^{+  0.20}_{-  0.21}$&$ 2690^{+  370}_{-  330}$&3\\\\
    OY Car&   90.8928&$0.840^{+0.040}_{-0.040}$&$0.1020^{+0.0030}_{-0.0030}$&$15000^{+ 2000}_{- 2000}$&$0.086^{+0.005}_{-0.005}$&$0.133^{+0.003}_{-0.003}$&$-10.10^{+  0.30}_{-  0.33}$&$ 2550^{+  500}_{-  450}$&1,2,3\\\\
    HT Cas&  106.0560&$0.610^{+0.040}_{-0.040}$&$0.1500^{+0.0100}_{-0.0100}$&$14000^{+ 1000}_{- 1000}$&$0.091^{+0.020}_{-0.020}$&$0.150^{+0.013}_{-0.013}$&$ -9.66^{+  0.24}_{-  0.24}$&$ 3000^{+  470}_{-  400}$&1,2\\\\
    DV UMa&  128.2800&$1.041^{+0.024}_{-0.024}$&$0.1510^{+0.0010}_{-0.0010}$&$20000^{+ 1500}_{- 1500}$&$0.157^{+0.010}_{-0.010}$&$0.204^{+0.016}_{-0.016}$&$ -9.97^{+  0.17}_{-  0.18}$&$ 2380^{+  470}_{-  300}$&1,2\\\\

 \enddata
\tablecomments{$P$ is the orbital period in minutes; $M_1$ and $M_2$ are the masses of the WD primary and the donor in the solar units, respectively; $q=M_2/M_1$ is the mass ratio; $T_1$ and $T_2$ are the effective temperatures of the WD and the donor in degrees Kelvin, respectively; $R_2$ is the radius of the donor in the solar units; $\dM$ is the mass-loss rate in units of $\dMd$; References: (1) \citet{c_Patterson_2005}; (2) \citet{c_Townsley_2009}; (3) \citet{c_Littlefair_2008}.}
\label{ta_ecl}
\end{deluxetable}


\begin{thebibliography} {}
\bibitem[Andronov et al.(2003)]{c_Andronov_2003}   Andronov, N., Pinsonneault, M., Sills, A. 2003, \apj, 582, 358A
\bibitem[Baraffe et al.(1998)]{c_Baraffe_1998} Baraffe, I., Chabrier, G., Allard, F., \& Hauschildt, P.\ H.\ 1998, \aap, 337, 403
\bibitem[Baraffe et al.(2002)]{c_Baraffe_2002} Baraffe, I., Chabrier, G., Allard, F., \& Hauschildt, P.\ H.\ 2002, \aap, 382, 563
\bibitem[Baraffe et al.(2003)]{c_Baraffe_2003} Baraffe, I., Chabrier, G., Barman, T.\ S., Allard, F., \& Hauschildt, P.\ H.\ 2003, \aap, 402, 701
\bibitem[Barker (2003)]{c_Barker_2003} Barker, J., \& Kolb, U.\ 2003, \mnras, 340, 623
%\bibitem[Beuermann et.\ al.(1998)]{c_Beuermann_1998} Beuermann, K., Baraffe, I., Kolb, U., \& Weichhold, M.\ 1998, \aap, 339, 518
\bibitem[Burrows \& Leibert(1993)]{c_Burrows_1993} Burrows, A., \& Liebert, J.\ 1993, RevModPhys, 65, 301
%\bibitem[Claret(1999)]{c_Claret_1999}Claret, A.\ 1999, \aap, 350, 56
%\bibitem[Chabrier \& Baraffe(1997)]{c_Chabrier_1997}Chabrier, G., \& Baraffe, I.\ 1997, \aap, 327, 1039
\bibitem[Chandrasekhar(1933a)]{cha33a} Chandrasekhar, S.\ 1933, \mnras, 93, 390
\bibitem[Chandrasekhar(1939)]{c_Chandra_1939}Chandrasekhar, S.\ 1939, An introduction to the study of stellar structure (Chicago Univ.\ Press)
\bibitem[D'Antona et.\ al.(1989)]{c_DAntona_1989}D'Antona, F., Mazzitelli, I., Ritter, H.\ 1989, \aap, 225, 391
\bibitem[Eggleton(1983)]{c_Eggleton_1983}Eggleton, P.\ P.\ 1983, \apj, 268, 368
\bibitem[Hansen \& Kavaler(1995)]{c_Hansen_1995}Hansen, C.\ J., \& Kawaler, S., D.\ 1995, Stellar interiors Physical principles, Structure, and Evolution, ed.\ Appenzeller, I., Harvit, M., Kippenhahn, R., Strittmatter, P., A., (Springel-Verlag, New-York, Inc.)
\bibitem[Howell et al.(2001)]{c_Howell_2001}    Howell, S.\ B., Nelson, L.\ A., \& Rappaport, S.\ 2001, \apj, 550, 897
\bibitem[King(1988)]{c_King_1988}King, A.\ R.   1988, \qjras, 29, 1
\bibitem[Kippenhahn \& Weigert(1990)]{c_Kipp_1990}Kippenhahn, R., Trimble, V., \& Weigert, A.\ 1990, Stellar Structure and Evolution, ed.\ Harvit, M., Kippenhahn, R., Trimble, V., Zahn, J.-P., (Springel-Verlag, Berlin Heidelberg)
\bibitem[Knigge(2006)]{c_Knigge_2006}Knigge, C.\ 2006, \mnras, 373, 484
\bibitem[Kolb \& Baraffe(1999)]{c_Kolb_1999}Kolb, U., \& Baraffe, I.\ 1999, \mnras, 309, 1034
%\bibitem[Kolb \& Ritter(1992)]{c_Kolb_1992}Kolb, U., \& Ritter, H.\ 1992, \aap, 254, 213
\bibitem[Kopal(1972)]{c_Kopal_1972}Kopal, Z.\ 1972, Adv.\ Astron.\ Ap.\ 9, 1
\bibitem[Landau \& Lifshitz(1975)]{c_Landau_1975}Landau, L.\ D., \& Lifshitz, E.\ M.\ 1975, The Classical Theory of Fields, 4th edn. (Pergamon, New York)
%\bibitem[Lauterborn \& Weigert(1972)]{c_Lauterborn_1972}Lauterborn, D., \& Weigert, A 1972, \aap, 18, 249.\
\bibitem[Littlefair et al.(2008)]{c_Littlefair_2008}Littlefair, S.\ P., Dhillon, V.\ S., Marsh, T.\ R., G\"{a}nsicke, B.\ T., Baraffe, I., \& Watson, C.\ A.\ 2008, \mnras, 388, 1582

\bibitem[Littlefair et al.(2007)]{c_Littlefair_2007}Littlefair, S.\ P., Dhillon, V.\ S., Marsh, T.\ R., G\"{a}nsicke, B.\ T., Baraffe, I., \& Watson, C.\ A.\ 2007, \mnras, 381, 827
%\bibitem[Nelson et.\ al.(1985)]{c_Nelson_1985}Nelson, L.\ A., Chau, W.\ Y., \& Rosenblum, A.\ 1985, \apj, 299, 658
%\bibitem[Paczynski(1971)]{c_Pacz_1971}Paczynski, B.\ 1971, \araa., 9, 183
\bibitem[Paczynski(1981)]{c_Pacz_1981}Paczynski, B.\ 1981, Acta Astron., 31, 1
%\bibitem[Paczynski \& Sienkiewicz(1983)]{c_Pacz_1983}Paczynski, B., \& Sienkiewicz, R.\ 1983, \apj, 268, 825
\bibitem[Patterson(1984)]{c_Patterson_1984}Patterson, J., et al.\ 1984, \apjs, 54, 443
\bibitem[Patterson et al.(2005)]{c_Patterson_2005}Patterson, J., et al.\ 2005, \pasp, 117, 1204
\bibitem[Patterson(2009)]{c_Patterson_2008}Patterson, J.\ 2009, \mnras, submitted (arXiv:0903.1006)
\bibitem[Patterson et al.(2008)]{c_Patterson_2008_k}Patterson, J., Thorstensen, J.\ R., \& Knigge, C. 2008, \pasp, 120, 510
%\bibitem[Politano(1996)]{c_Politano_1996}Politano, M.\ 1996, \apj, 465, 338
\bibitem[Rappaport et al.(1982)]{c_Rappaport_1982}Rappaport, S., Joss, P.\ C., \& Webbink, R.\ F.\ 1982, \apj, 254, 616
%\bibitem[Ribas(2006)]{c_Ribas_2006}Ribas, I.\ 2006, \aaps, 304, 89
\bibitem[Renvoiz\'{e} et al.(2002)]{c_Renvoize_2002}Renvoiz\'{e}, V., Baraffe, I., Kolb, U., \& Ritter, H.\ 2002, \aap, 389, 485
\bibitem[Rezzolla et.\ al.(2001)]{c_Rezzolla_2001}Rezzolla, L., Uryu, K., \& Yoshida, S.\ 2001, \mnras, 888
%\bibitem[Ritter (1988)]{c_Ritter_1988}Ritter, H.\ 1988, \aap, 202, 93
\bibitem[Ritter(1988)]{c_Ritter_1988} Ritter, H.\ 1988, \aap, 202,93
\bibitem[Ritter(1996)]{c_Ritter_1996}Ritter, H. 1996, in Evolutionary Processes in Binary Stars, Proceedings of the NATO Advanced Study Institute, Cambridge, ed. Wijers, R., \& Davies, M., 477, 223
\bibitem[Ritter \& Kolb(2003)]{c_Ritter_2003}Ritter, H., \& Kolb, U. 2003, \aap, 404, 301
\bibitem[Sirotkin \& Kim(2009)]{c_SWTK_2009} Sirtokin, F.\ V., \& Kim, W.-T.\ 2009, \apj, 698, 715(SK09)
\bibitem[Sion(1995)]{c_Sion_1995} Sion, E., M., 1995, \apj, 438, 876
\bibitem[Soberman et al.(1997)]{c_Soberman_1997} Soberman, G. E., Phinney, E.S., \& Heuvel E.P.J. 1997, \aap, 327, 620
\bibitem[Stehle et al.(1996)]{c_Stehle_1996}Stehle, R., Ritter, H., \& Kolb, U.\ 1996, \mnras, 279, 581
%\bibitem[Stothers(1974)]{sto74}Stothers, R., 1974, \apj, 194, 651
\bibitem[Townsley \& Bildsten(2003)]{c_Townsley_2003}Townsley D.\ M., \& Bildsten L. 2003, \apj, 596, L227
\bibitem[Townsley \& Bildsten(2004)]{c_Townsley_2004}Townsley D.\ M., \& Bildsten L. 2004, \apj, 600, 390
\bibitem[Townsley \& G{\aa}nsicke(2009)]{c_Townsley_2009}Townsley, D.\ M., \& G{\aa}nsicke, B.\ T. 2009, \apj, 693, 1007
\bibitem[G{\aa}nsicke et al.(2009)]{c_Gansike_2009}G{\aa}nsicke, B.\ T. et al.\ 2009, \mnras, 397, 2170
\bibitem[Uryu \& Eriguchi(1999)]{c_Uryu_1999}Uryu, K., \& Eriguchi, Y.\ 1999, \mnras, 303, 329
\bibitem[Urban \& Sion(2006)]{c_Urban_2006} Urban, J. A, \& Sion E. M. 2006, \apj, 642, 1029
%\bibitem[Meyer \& Meyer-Hofmeister(1983)]{c_Meyer_1983}Meyer, F., \& Meyer-Hofmeister, E.\ 1983, \aap, 121, 29
%\bibitem[Mochnacki(1984)]{c_Moch_1984}Mochnacki, S.\ W.\ 1984, \apjs, 55, 551
%\bibitem[Willems at.\ al.(2005)]{c_Willems_2005}Willems B., Kolb U., Sandquist E. L., \& Taam R. E., Dubus G., 2005, \apj, 635, 1263
%\bibitem[Weisberg \& Taylor(2005)]{c_Weisberg_2005}Weisberg, J.\ M., \& Taylor, J.\ H. 2005, ASPC, 328, 25
\end{thebibliography}
\end{document}